\pdfoutput=1

\documentclass[11pt,twoside,a4paper,cmspaper,final,collab]{cms-tdr}
\def\svnVersion{343677}\def\cmsCernNoTag{CERN-EP/2016-062}\def\cmsCernDate{\today}\def\cmsMessage{Published in Physical Review D as \href{http://dx.doi.org/10.1103/PhysRevD.93.092006}{\doi{10.1103/PhysRevD.93.092006.}}}

\begin{document}\cmsNoteHeader{TOP-12-030}

\hyphenation{had-ron-i-za-tion}
\hyphenation{cal-or-i-me-ter}
\hyphenation{de-vices}
\RCS$Revision: 339795 $
\RCS$HeadURL: svn+ssh://svn.cern.ch/reps/tdr2/papers/TOP-12-030/trunk/TOP-12-030.tex $
\RCS$Id: TOP-12-030.tex 339795 2016-04-22 15:23:42Z stiegerb $
\newlength\cmsFigWidth
\ifthenelse{\boolean{cms@external}}{\setlength\cmsFigWidth{0.85\columnwidth}}{\setlength\cmsFigWidth{0.4\textwidth}}
\ifthenelse{\boolean{cms@external}}{\providecommand{\cmsLeft}{top}}{\providecommand{\cmsLeft}{left}}
\ifthenelse{\boolean{cms@external}}{\providecommand{\cmsRight}{bottom}}{\providecommand{\cmsRight}{right}}
\providecommand{\mtop}{\ensuremath{m_{\cPqt}}}
\providecommand{\lxy}{{$L_{\rm xy}$}}
\providecommand{\lxyhat}{{$\widehat{L_{\rm xy}}$}}
\providecommand{\MET}{\ensuremath{E_{\rm T}^{\rm miss}}}
\providecommand{\ttV}{\ensuremath{\ttbar+\PV}\xspace}
\newcommand{\msvl}{\ensuremath{m_{\mathrm{svl}}}\xspace}
\newcommand{\muR}{\ensuremath{\mu_{\mathrm{R}}}\xspace}
\newcommand{\muF}{\ensuremath{\mu_{\mathrm{F}}}\xspace}
\newcommand{\ee}{\ensuremath{\Pe\Pe}\xspace}
\newcommand{\mumu}{\ensuremath{\Pgm\Pgm}\xspace}
\newcommand{\emu}{\ensuremath{\Pe\Pgm}\xspace}
\newcommand{\ejets}{\ensuremath{\Pe}\xspace}
\newcommand{\mujets}{\ensuremath{\Pgm}\xspace}
\newcommand{\ztwostar}{{\rm Z2*}\xspace}
\newcommand{\ztsrblep}{\ztwostar\,LEP $r_{\PQb}$\xspace}
\newcommand{\PDstp}{\ensuremath{{\PD^{\ast}{(2010)}^{+}}}}

\cmsNoteHeader{TOP-12-030}
\title{Measurement of the top quark mass using charged particles \texorpdfstring{in $\Pp\Pp$ collisions at $\sqrt{s}=8\TeV$}{in pp collisions at sqrt(s) = 8 TeV}}

\date{\today}

\abstract{
A novel technique for measuring the mass of the top quark that uses only the kinematic properties of its charged decay products is presented. Top quark pair events with final states with one or two charged leptons and hadronic jets are selected from the data set of 8\TeV proton-proton collisions, corresponding to an integrated luminosity of 19.7\fbinv. By reconstructing secondary vertices inside the selected jets and computing the invariant mass of the system formed by the secondary vertex and an isolated lepton, an observable is constructed that is sensitive to the top quark mass that is expected to be robust against the energy scale of hadronic jets. The main theoretical systematic uncertainties, concerning the modeling of the fragmentation and hadronization of $\PQb$~quarks and the reconstruction of secondary vertices from the decays of $\PQb$~hadrons, are studied. A top quark mass of $173.68\pm 0.20\stat ^{+1.58}_{-0.97}\syst\GeV$ is measured. The overall systematic uncertainty is dominated by the uncertainty in the \PQb~quark fragmentation and the modeling of kinematic properties of the top quark.
}

\hypersetup{%
pdfauthor={CMS Collaboration},%
pdftitle={A measurement of the top quark mass using charged particles},%
pdfsubject={CMS},%
pdfkeywords={CMS, physics, top, mass, tracking, b-tagging}}

\maketitle

\section{Introduction}\label{sec:intro}
The top quark is the heaviest known elementary particle and as such has a privileged interaction with the Higgs boson.
Its mass, \mtop, is hence an important input to global fits of electroweak parameters together with measurements of the \PW~boson and Higgs~boson masses, and serves as an important cross-check of the consistency of the standard model (SM).
Moreover, by comparing precision electroweak measurements and theoretical predictions, a precisely measured \mtop{} can place strong constraints on contributions from physics beyond the SM.\@
The top quark is the only colored particle that decays before forming a color-neutral state through hadronization and thus presents a unique opportunity to directly probe the properties of color charges.

Direct determinations of the mass of the top quark have been carried out with ever-increasing precision since it was discovered at the Tevatron by the CDF and D0 experiments~\cite{Abe:1995hr,Abachi:1995iq}.
More recently, the most precise measurements reconstruct top quarks in hadronic decays and calibrate the energy of hadronic jets in-situ, using constraints from the reconstructed \PW{}~boson mass~\cite{Chatrchyan:2012cz,Chatrchyan:2013xza,Aad:2015nba}.
Other analyses exploit the purity of leptonic top quark decays and constrain the neutrino momenta analytically~\cite{Chatrchyan:2012ea,Aad:2015nba}.
All four experiments where the top quark mass is being studied (ATLAS, CDF, CMS, and D0) have combined their results in a world average~\cite{worldcomb}.
A recent combination of measurements at 7 and 8\TeV\ by the CMS experiment yields the best determination of the top quark mass to date, with a result of $172.44\pm0.48\GeV$, \ie\ reaching a precision of $0.28\%$~\cite{Khachatryan:2015hba}.

The most precise top quark mass measurements are systematically limited by experimental uncertainties related to the calibration of reconstructed jet energies and their resolution, with other important uncertainties concerning the modeling of the fragmentation and hadronization of bottom quarks.
To improve further the precision of the value of the top quark mass and our understanding of the modeling of top quark decays, the development and application of alternative and complementary methods is essential.
Complementarity to ``standard'' methods can be gained by using observables with reduced sensitivity to certain sources of systematic uncertainties, such as the $\PQb$~hadron decay length~\cite{Hill:2005zy,Abulencia:2006rz,Aaltonen:2009hd} or kinematic properties of leptons~\cite{Frixione:2014ala}, or by extracting the mass from endpoints of kinematic distributions~\cite{Chatrchyan:2013boa} or from the production cross section~\cite{Khachatryan:2016mqs}.

This paper describes a measurement performed with the CMS experiment at the CERN LHC that minimizes the sensitivity to experimental systematic uncertainties such as jet energy scale.
This is achieved by constructing a mass-dependent observable that uses only the individually-measured momenta of charged decay products (tracks) of the top quark.
The mass of the top quark is estimated by measuring the invariant mass of a charged lepton from the \PW~boson decay and the tracks used in the reconstruction of a secondary vertex (SV) resulting from the long lifetime of $\PQb$~hadrons.
The dependence of the observable on the top quark mass is calibrated using simulated Monte Carlo (MC) events.
This approach is similar to a proposed measurement using the invariant mass of leptons and reconstructed \PJGy{} mesons~\cite{Kharchilava:1999yj}, but requires a lower integrated luminosity to become sensitive.

The paper is organized as follows: Section~\ref{sec:experiment} describes the experiment, the collected and simulated data, and the event reconstruction and selection; Section~\ref{sec:modeling} describes control region studies of \cPqb\ quark fragmentation and secondary vertex reconstruction; Section~\ref{sec:topmass} describes the measurement of the top quark mass and the assigned systematic uncertainties; and Section~\ref{sec:conclusions} concludes and gives an outlook of prospects in the ongoing LHC run.\@

\section{Experimental setup}\label{sec:experiment}

\subsection{The CMS detector}
The central feature of the CMS apparatus is a superconducting solenoid of 6~m internal diameter, providing a magnetic field of 3.8~T.
Within the solenoid volume are a silicon pixel and strip tracker, a lead tungstate crystal electromagnetic calorimeter (ECAL), and a brass and scintillator hadron calorimeter (HCAL), each composed of a barrel and two endcap sections.
The tracker has a track-finding efficiency of more than 99\% for muons with transverse momentum $\pt > 1\GeV$ and pseudorapidity $|\eta| < 2.5$.
The ECAL is a fine-grained hermetic calorimeter with quasi-projective geometry, and is segmented in the barrel region of $|\eta| < 1.48$ and in two endcaps that extend up to $|\eta| < 3.0$.
The HCAL barrel and endcaps similarly cover the region $|\eta| < 3.0$.
In addition to the barrel and endcap detectors, CMS has extensive forward calorimetry.
Muons are measured in gas-ionization detectors which are embedded in the  flux-return yoke outside of the solenoid.
A more detailed description of the CMS detector, together with a definition of the coordinate system used and the relevant kinematic variables, can be found in Ref.~\cite{Chatrchyan:2008zzk}.

\subsection{Data and simulation}
This analysis makes use of a large sample of top quark pair, \ttbar, event candidates with either one or two isolated charged leptons (electrons or muons) in the final state.
In the semileptonic (only one lepton) case,
at least four reconstructed hadronic jets are required,
whereas in the dilepton case at least two jets are required.
Events are selected from the data sample acquired in proton-proton ($ \Pp \Pp $) collisions at a center-of-mass energy of $\sqrt{s}=8\TeV$ by the CMS experiment throughout 2012, corresponding to an integrated luminosity of 19.7\fbinv.

{\sloppy
At that energy the predicted \ttbar\ cross section in $ \Pp \Pp $ collisions,
computed at the next-to-next-to-leading-order (NNLO) quantum chromodynamics (QCD) and including corrections and next-to-next-to-leading-logarithmic resummation accuracy~\cite{Czakon:2013goa}, is
\mbox{$245.8^{+8.7}_{-10.6}~{\rm pb}$}
for a top quark mass of $173\GeV$,
where the uncertainty covers missing higher orders in the calculation as well as variations of the parton distribution functions (PDFs).
Signal \ttbar\ events are simulated with the leading-order (LO) \MADGRAPH\ (v5.1.3.30) generator~\cite{Alwall:2014hca} matched to LO \PYTHIA\ (v6.426)~\cite{Sjostrand:2006za} for parton showering and fragmentation.
The $\tau$ lepton decays are simulated with the \TAUOLA\ package (v27.121.5)~\cite{Was:2000st}.
The LO CTEQ6L1 PDF set~\cite{Pumplin:2002vw} and the \ztwostar\ underlying event tune~\cite{Chatrchyan:2011id} are used in the generation.
The \ztwostar\ tune is derived from the Z1 tune~\cite{Field:2010bc}, which uses the CTEQ5L PDF set, whereas \ztwostar\ adopts CTEQ6L.
Matrix elements describing up to three partons in addition to the \ttbar\ pair are included in the generator used to produce the simulated signal samples, and the MLM prescription~\cite{Mangano:2006rw} is used for matching of matrix-element jets to parton showers.
Following the generator chain, the response of the CMS detector
is simulated using \GEANTfour\ (v.9.4p03) for both signal and background samples~\cite{Agostinelli:2002hh}.
}

The most relevant background for the semileptonic channel is the production of a \PW~boson in association with hadronic jets.
This background is modeled with \MADGRAPH\ and normalized to a total cross section of 36.3\unit{nb}, computed with \textsc{fewz} (v3.1)~\cite{Gavin:2012sy} at NNLO.\@
Multijet QCD processes are also relevant and studied in simulations using \PYTHIA.\@
Single top quark processes are modeled with \POWHEG\ (v1.0, r1380)~\cite{Nason:2004rx,Frixione:2007vw,Alioli:2010xd,Alioli:2009je,Re:2010bp} with the CTEQ6M PDF set and normalized to the cross sections of 22.2, 86.1, and 5.6\unit{pb} for the $\cPqt\PW$, $t$, and $s$ production channels, respectively~\cite{Kidonakis:2012eq}.
Charged-lepton production from Drell--Yan (DY) processes is modeled with \MADGRAPH\ for dilepton invariant masses above 10\GeV\ and is normalized to a cross section of 4.4\unit{nb},
computed with \textsc{fewz}~\cite{Gavin:2010az,Li:2012wna}.
The production of $\PW \PW$, $\PW \PZ$, and $\PZ \PZ$ pairs is modeled with \PYTHIA\ and normalized to cross sections of 54.8, 33.2, and 17.7\unit{pb}, respectively, computed at next-to-leading order (NLO) accuracy using \MCFM\ (v6.6)~\cite{Campbell:2010ff}.

All simulated samples include the effects of pileup, \ie\ multiple $ \Pp \Pp $ collisions in the same and neighboring beam crossings (within 50\unit{ns}) as the generated hard interaction.
The distribution of the number of pileup events in simulation matches that in the data and has an average of about 21 interactions per bunch crossing.

\subsection{Event reconstruction and selection}\label{sec:eventsel}
The event selection is designed to identify the \ttbar\ final state in the semileptonic and dileptonic channels.
Single-lepton triggers are used to collect the data samples for the semileptonic channels with a minimum \pt\ of 27 for electrons and 24\GeV for muons.
In the dilepton channel double-lepton triggers are required with a minimum \pt\ of 17 and 8\GeV for the leading and sub-leading leptons, respectively.
In both cases isolation and identification criteria are required at the trigger level.
More information can be found in Refs.~\cite{Khachatryan:2016mqs,Khachatryan:2016yzq}.

The events are reconstructed using a particle-flow (PF) algorithm that optimally combines the information from all subdetectors
to reconstruct and identify all individual particles in the event~\cite{CMS-PAS-PFT-09-001,CMS-PAS-PFT-10-001}.
In addition, improved electron and muon reconstruction, identification and calibration algorithms have been employed as described in~\cite{Khachatryan:2015hwa,Chatrchyan:2013sba}.
Electron candidates are required to have $\pt>30\GeV$ and to be in the fiducial region of the detector, \ie\ $\vert\eta\vert\leq2.4$.
Muon candidates are selected with $\pt>26\GeV$ and $\vert\eta\vert\leq2.1$.

In the dilepton channel these requirements are relaxed to $\pt>20\GeV$ and $\vert\eta\vert\leq 2.4$ for all lepton candidates.
The track associated with each lepton candidate is required to have an impact parameter compatible with prompt production.
A particle-based relative isolation is computed for each lepton and is corrected on an event-by-event basis for contributions from pileup events~\cite{Khachatryan:2016mqs}.
The scalar sum of the transverse momenta of all reconstructed particle candidates---except for the leptons themselves---within a cone of size $\Delta R=\sqrt{\smash[b]{(\Delta \eta)^{2}+(\Delta \phi)^{2}}}<0.3$ ($<0.4$ for muons) built around the lepton direction must be less than 10\% of the electron \pt\ and less than 12\% of the muon \pt.
In the dilepton channels, the electron isolation threshold is relaxed to less than $15\%$.
Events in the semileptonic channel are required to have exactly one selected lepton, with a veto on additional leptons.
In the dilepton channel, at least two selected leptons are required.

Jets are reconstructed using the anti-\kt algorithm with a distance parameter of $0.5$ and taking PF candidates as input to the clustering~\cite{Cacciari:2008gp}.
The jet momentum is defined as the vectorial sum of all particle momenta associated to the jet and is determined from the simulation to be within 5--10\% of the generated jet momentum at particle level over the whole \pt\ range and detector acceptance.
An offset correction is applied to take into account the extra energy clustered into the jets due to pileup, following the procedure described in Refs.~\cite{Cacciari:2008gn,Cacciari:2007fd}.
Jet energy scale corrections are derived from the simulation and are cross-checked with in-situ measurements of the energy balance in dijet and photon+jet events.
The selected jets are required to have a corrected \pt\ greater than $30\GeV$ and $\vert\eta\vert\leq 2.5$.
Jets within $\Delta R = 0.4$ of any selected lepton are rejected, but the event is retained if it passes the other selection criteria.
The magnitude of the vectorial sum of the transverse momenta of all PF candidates reconstructed in the event is used as an estimator of the energy imbalance in the transverse plane, \MET.\@

For each jet, the charged PF candidates used in the clustering are given as input to an adaptive vertex fitter algorithm to reconstruct secondary vertices~\cite{Fruhwirth:2007hz}.
Secondary vertex candidates that share $65\%$ or more of their tracks with the primary vertex (defined as the vertex with highest $\sum{\pt^2}$ of its associated tracks) or that have a flight direction outside a $\Delta R=0.5$ cone around the jet momentum are rejected.
Furthermore, if the radial distance from the primary vertex is greater than 2.5~cm, candidates with an invariant mass consistent with that of a \PKz, or higher than 6.5\GeV, are rejected (assuming each decay particle to have the rest mass of a charged \Pgp).
In case an event does not have any jet with a valid secondary vertex candidate it is discarded from the analysis.

Secondary vertices are used together with track-based lifetime information in a likelihood ratio algorithm to provide a discriminant for jets originating from the hadronization of a \cPqb~quark (``\cPqb\ jets'')~\cite{Chatrchyan:2012jua}.
The chosen threshold on the discriminant output value has an efficiency for selecting a genuine \cPqb~jet of about 60\%, selects charm-initiated jets with an efficiency of about 15\%, while the probability to misidentify a light-flavor jet as a \cPqb\ jet is about 1.5\%.
Jets passing this selection are referred to as \cPqb-tagged.

Events in the three dilepton channels (\emu, \ee, and \mumu) are selected with at least two jets, of which at least one is required to have a reconstructed secondary vertex.
The dilepton invariant mass is required to be greater than $20\GeV$ to remove low-mass QCD resonances.
To suppress contributions from DY production in the \ee\ and \mumu\ channels, the dilepton mass is further required to differ by at least $15\GeV$ from the \PZ~boson mass ($91\GeV$), and $\MET>40\GeV$ is required.
In the two semileptonic channels, events are selected with at least four jets, of which at least one has a reconstructed secondary vertex and one more has either another secondary vertex or is \cPqb-tagged.

\begin{figure}[htp]
\centering
\includegraphics[width=0.45\textwidth]{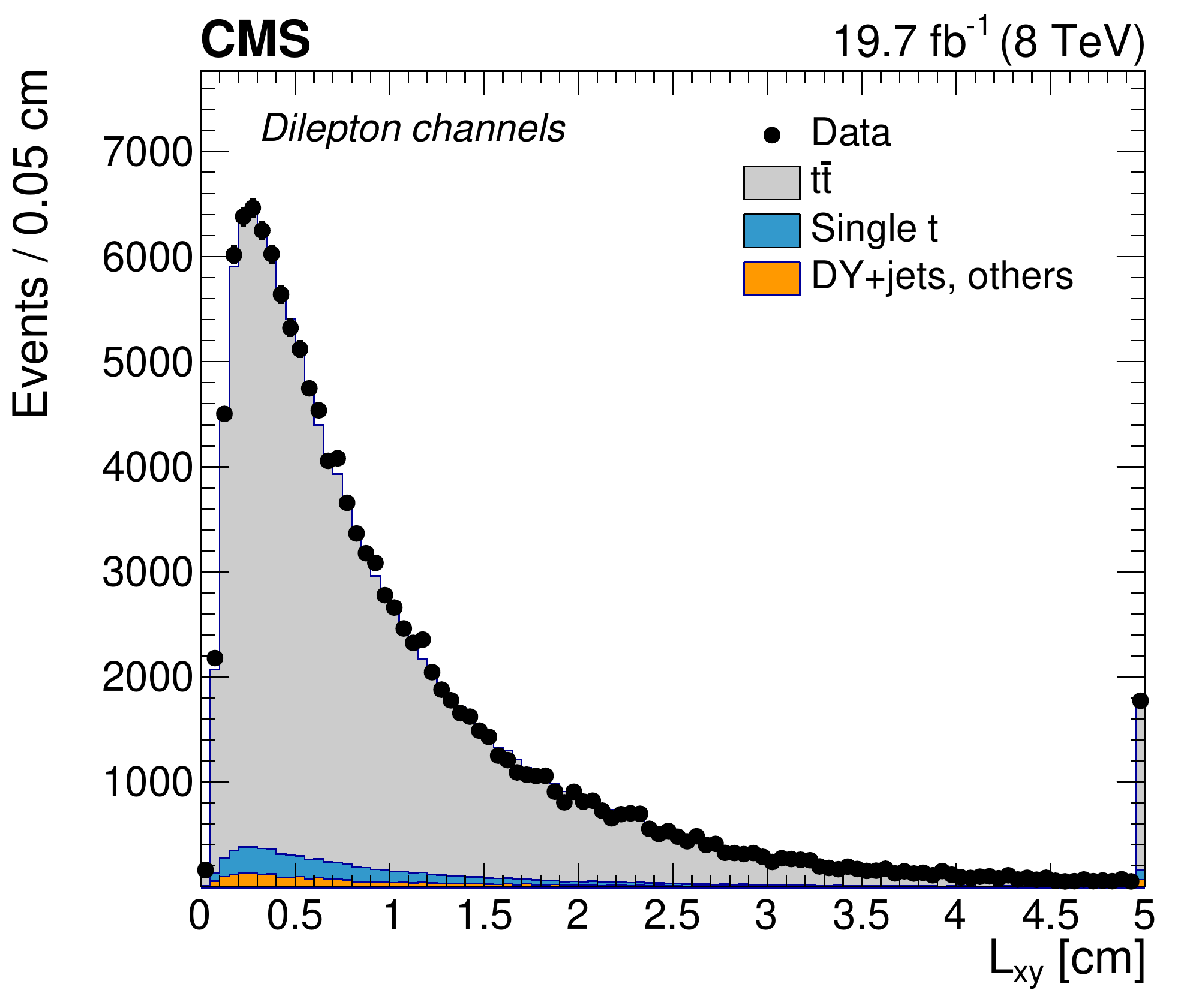}
\includegraphics[width=0.45\textwidth]{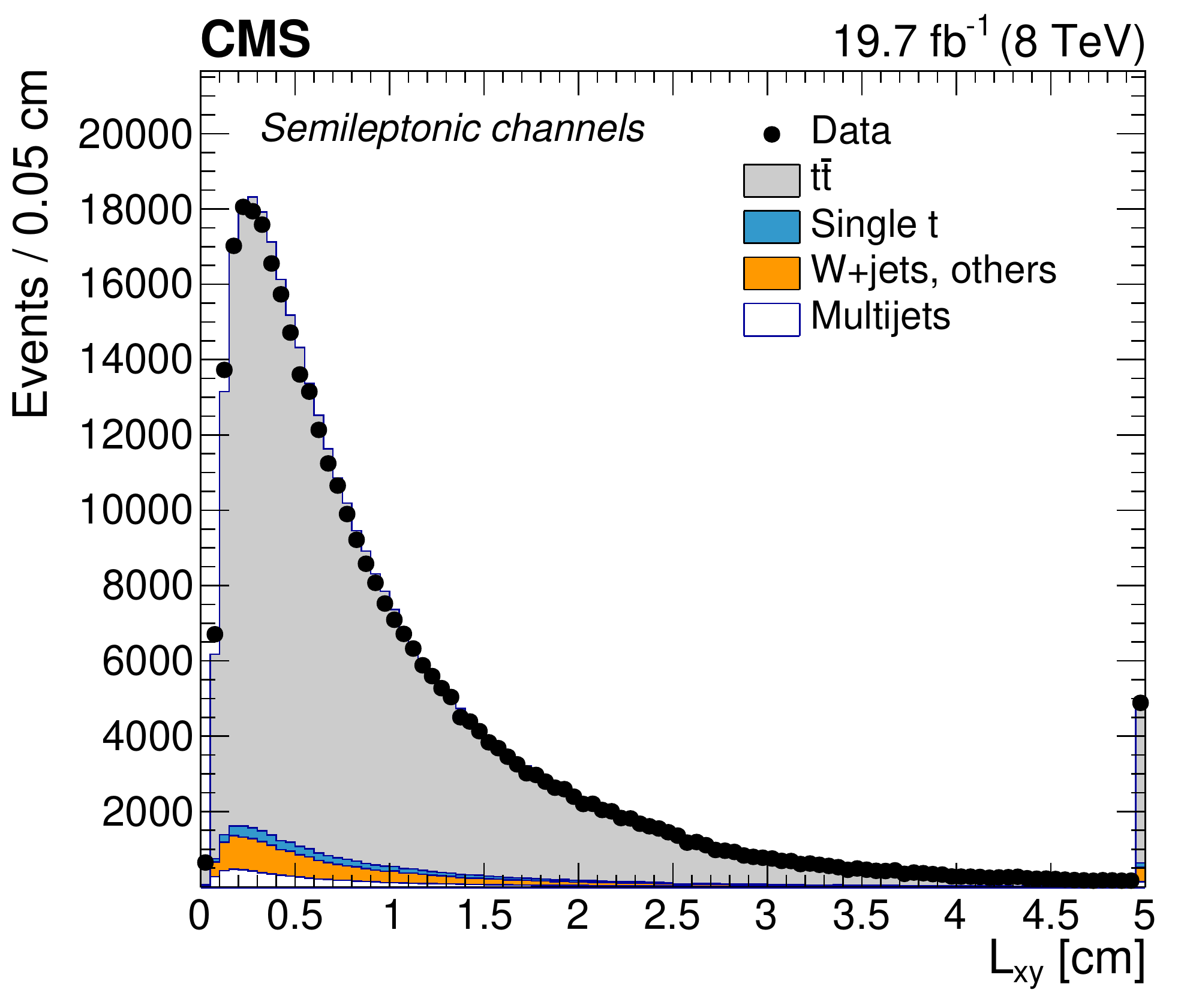}
\caption{
Distributions of the transverse decay length of secondary vertices with respect to the primary vertex in dilepton (\cmsLeft) and semileptonic channels (\cmsRight).
The expectations from simulation and estimates from the data for the multijet background are compared to the reconstructed data.
The last bin contains the overflow events.
}
\label{fig:lxy}
\end{figure}

Table~\ref{tab:eventyields} shows the number of selected data events in the five channels and the purity of events containing top quarks as expected from simulation.
Figure~\ref{fig:lxy} shows the distribution of the transverse decay length, \lxy, between the secondary vertex reconstructed from charged-particle tracks inside the jets selected for this analysis and the primary vertex of each event.
Good agreement is observed between data and expectations based on $\mtop = 172.5\GeV$.
The background expectations are obtained from the simulation, except for the multijet background which is determined from a control region in the data, as described in Section~\ref{ssec:signalmodel}.

\begin{table}[htp]
\centering
\topcaption{
Number of observed events and expected purity of top quark production (\ttbar\ and single top quarks) for the five channels investigated in this analysis.
}
\label{tab:eventyields}
\begin{scotch}{lccccc}
	                & \emu\  & \ee\   & \mumu\ & \ejets\ & \mujets\ \\
	\hline

	Observed events & 31\,639  & 9\,558   & 10\,674  & 103\,586  & 117\,198  \\
	Expected purity & 98.6\% & 95.8\% & 95.4\% & 93.7\%  & 92.8\%  \\
\end{scotch}
\end{table}

\section{Analysis of \texorpdfstring{\cPqb\ quark}{b quark} fragmentation in data}\label{sec:modeling}
The crucial objects used in this measurement are the charged leptons from a \PW{}~boson decay and the charged decay products of a $\PQb$~hadron, forming a reconstructed secondary vertex.
While the reconstruction of leptons is well-controlled in the experiment, the modeling of hadronization of the colored decay products of the top quark is subject to theoretical uncertainties.
These uncertainties affect the kinematic properties of the produced tracks, as well as their flavor composition and multiplicity.

The parton-to-hadron momentum transfer in the hadronization of \cPqb{} quarks---referred to in the following as \cPqb\ quark fragmentation---has been measured before in $\Pep\Pem$ collisions by the ALEPH, DELPHI, OPAL, and SLD Collaborations~\cite{Heister:2001jg,Abbiendi:2002vt,DELPHI:2011aa,Abe:1999ki,Abe:2002iq}
, and in \Pp{}\Pap{} collisions by the CDF Collaboration~\cite{Affolder:1999iq}.
However, no measurement at the LHC has been published so far.

In this section, two complementary studies are presented that attempt to constrain the uncertainties from the modeling of \cPqb\ quark fragmentation, which are expected to be the main contributors to the final uncertainty in this top quark mass measurement.
These studies constitute a first step towards measuring the \cPqb\ quark fragmentation using \ttbar{} events, but, as will become clear, the 2012 LHC data do not provide the necessary statistical precision, and significant constraints on the \cPqb\ quark fragmentation will be possible only with future data.

In this study we compare the \PYTHIA\ \ztwostar\ tune, used by the CMS experiment at 8\TeV~\cite{Chatrchyan:2011id} with an updated version which includes the $\Pep\Pem$ data to improve the description of the fragmentation.
Without the inclusion of this data, the default \ztwostar\ \cPqb\ quark fragmentation function is found to be too soft.
The $r_{\cPqb}$ parameter in \PYTHIA\ (\verb|PARJ(47)|) can be optimized to fit the $\Pep\Pem$ data using the \textsc{Professor} tool~\cite{Buckley:2009bj},
resulting in a value of $0.591^{+0.216}_{-0.274}$.
In contrast, the default central value used in \ztwostar\ is 1.0~\cite{Seidel:2015vla}.
In this analysis, the improved tune using the
$r_{\cPqb}$ central value of 0.591 (and variations within the uncertainty band) is denoted as \ztsrblep\ (\ztsrblep~$\!^{\pm}$) and is used to calibrate the measurement and evaluate the systematic uncertainty associated with the calibration.
For completeness, we also include other alternatives of the \ztwostar\ tune using the Peterson and Lund parameterizations~\cite{Sjostrand:2006za}.
All the considered \PYTHIA\ tunes use the so-called Lund string fragmentation model~\cite{Andersson:1983ia}.
The impact on the measurement of \mtop\ when using the alternative cluster model~\cite{Webber:1983if,Winter:2003tt} is discussed in Section~\ref{sssec:theosysts}.

\subsection{Secondary vertex properties \texorpdfstring{in $\PZ$+jets and \ttbar\ events}{in Z+jets and t-tbar events}}
Events with a leptonically-decaying \PZ\ boson recoiling against hadronic jets provide an independent and low-background sample to study the properties of secondary vertices.
Candidate \PZ\ events are selected by requiring two opposite-sign leptons with an invariant mass compatible with the \PZ{}~boson mass within 15\GeV.
To minimize effects from mismodeling of kinematic properties of the \PZ{}~boson, events are reweighted such that the predicted $\pt{}(\PZ)$ distribution reflects the one observed in the data.
Furthermore, events are required to have a leading jet with $\pt>30\GeV$ that is spatially separated from the \PZ~boson candidate by $\Delta R>2.1$.

The flavor of jets with reconstructed secondary vertices in such events changes with increasing number of tracks associated with the vertex.
From simulation, we expect vertices with two tracks to predominantly correspond to jets from light and \cPqc~quarks, with the fraction of jets from \cPqb{} quarks increasing to above 90\% for vertices with five or more tracks.

Several observables of secondary vertex kinematic properties are investigated for their sensitivity to modeling of \cPqb\ quark fragmentation.
Of those, the highest sensitivity is achieved when studying the ratio of SV transverse momentum---\ie\ the transverse component of the vectorial sum of all charged particle momenta used in the reconstruction of the vertex---to the total transverse momentum of the jet carried by charged particles,

\begin{equation*}
	\mathcal{F}_\mathrm {ch} = \frac{\pt(\mathrm{SV})}{\vert\sum_{\mathrm{ch}} \vec{\pt}\vert}.
\end{equation*}

\begin{figure*}[htp]
\centering
\includegraphics[width=0.32\textwidth]{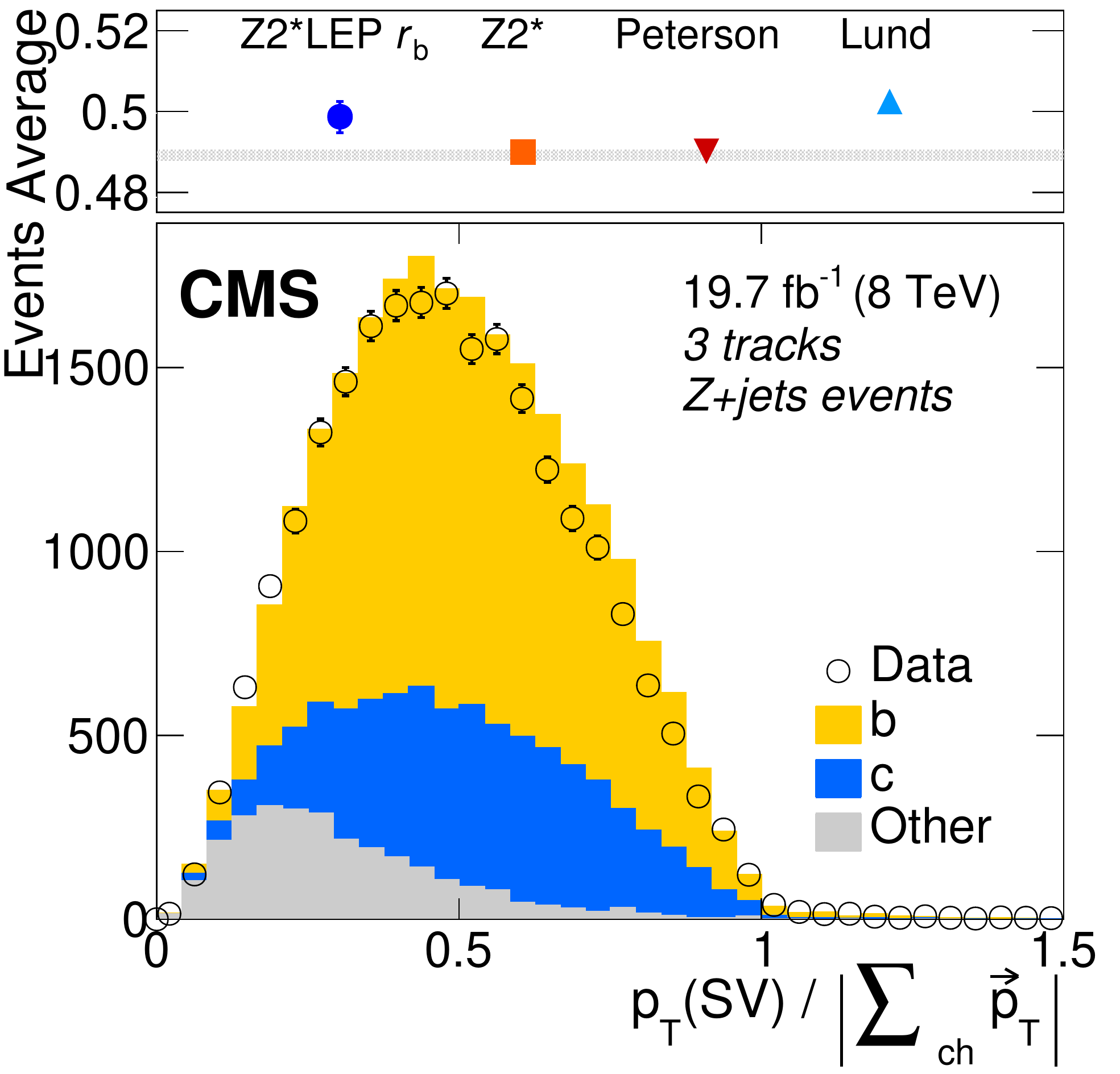}
\includegraphics[width=0.32\textwidth]{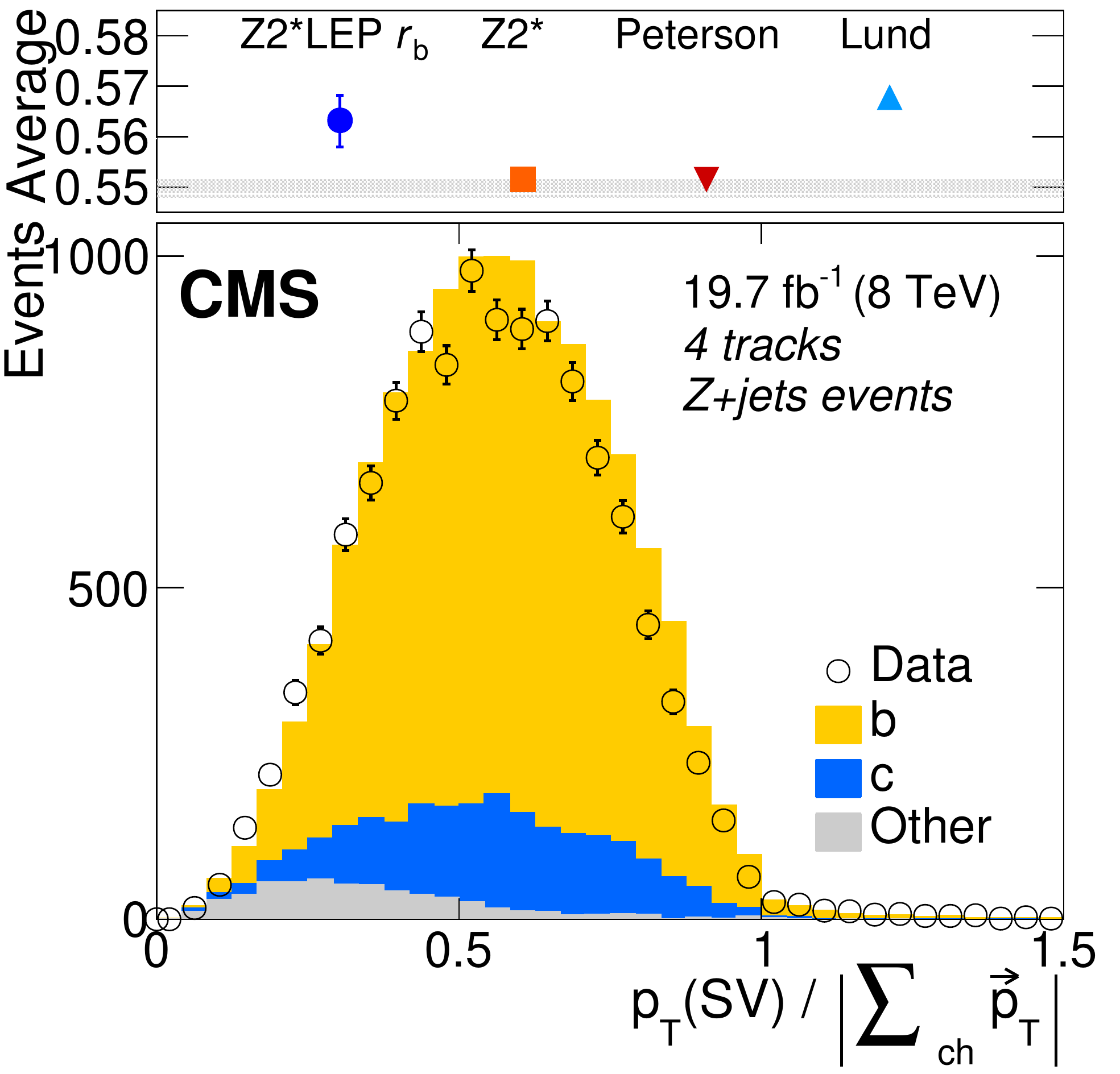}
\includegraphics[width=0.32\textwidth]{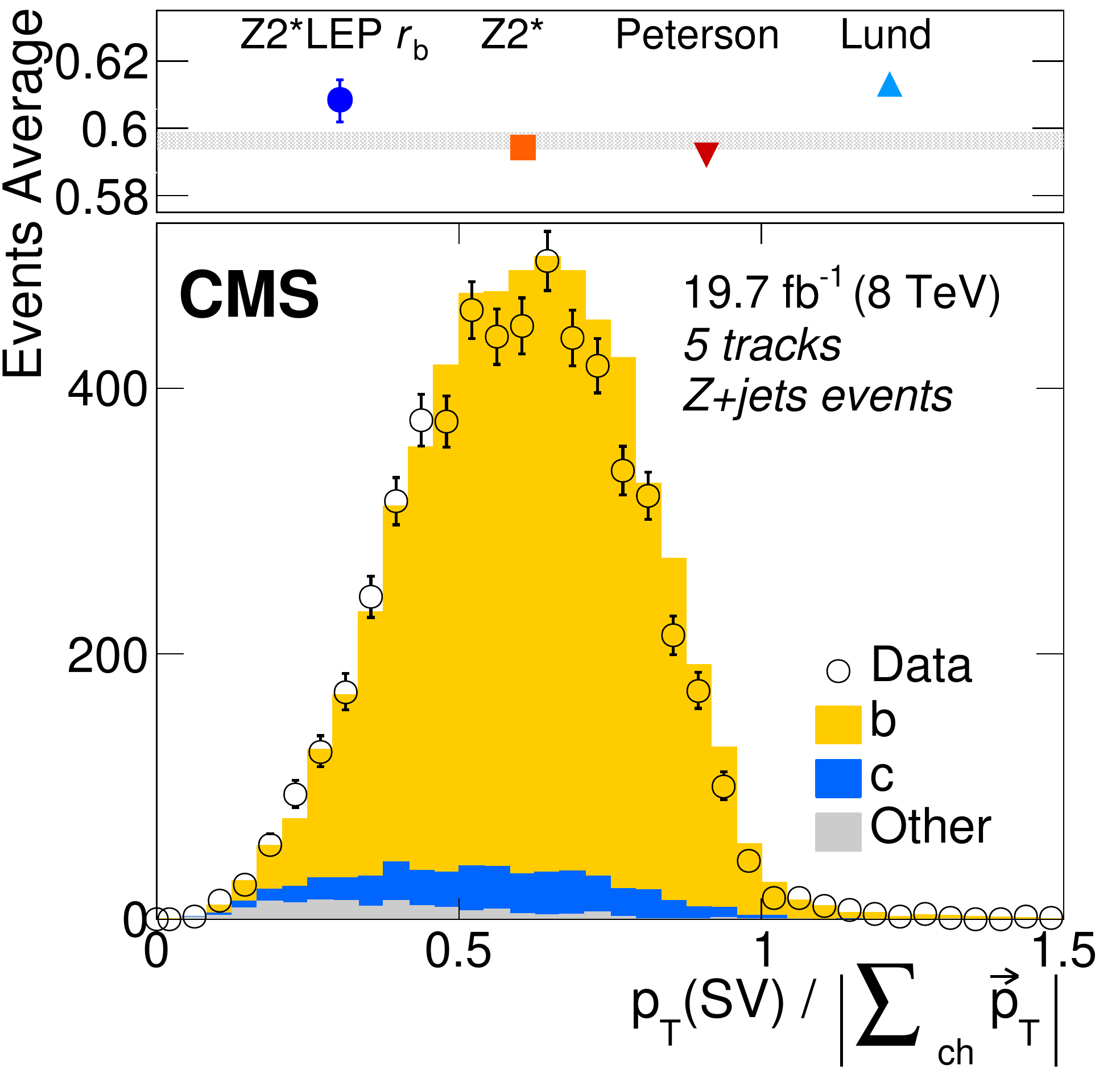}\\
\includegraphics[width=0.32\textwidth]{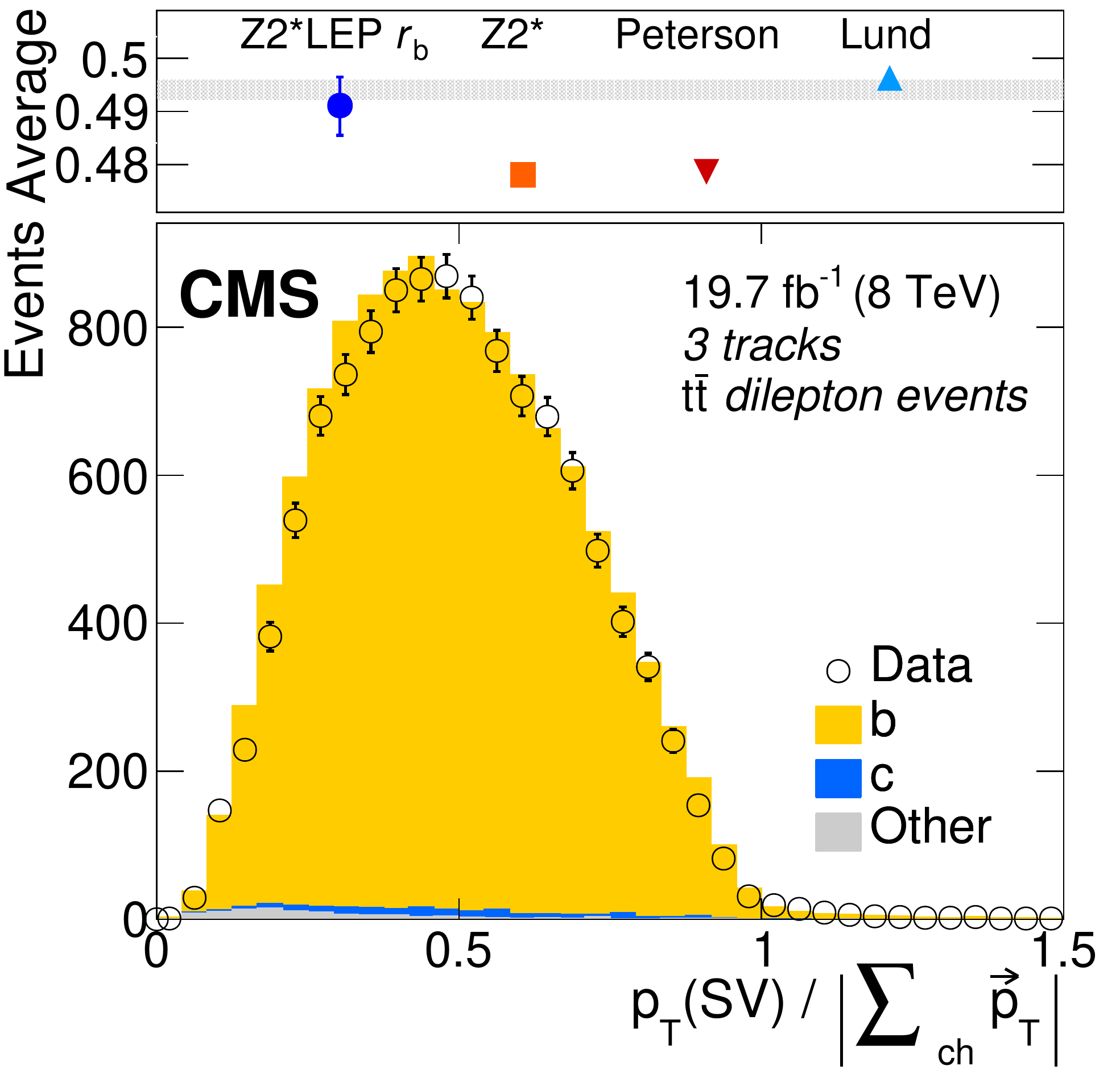}
\includegraphics[width=0.32\textwidth]{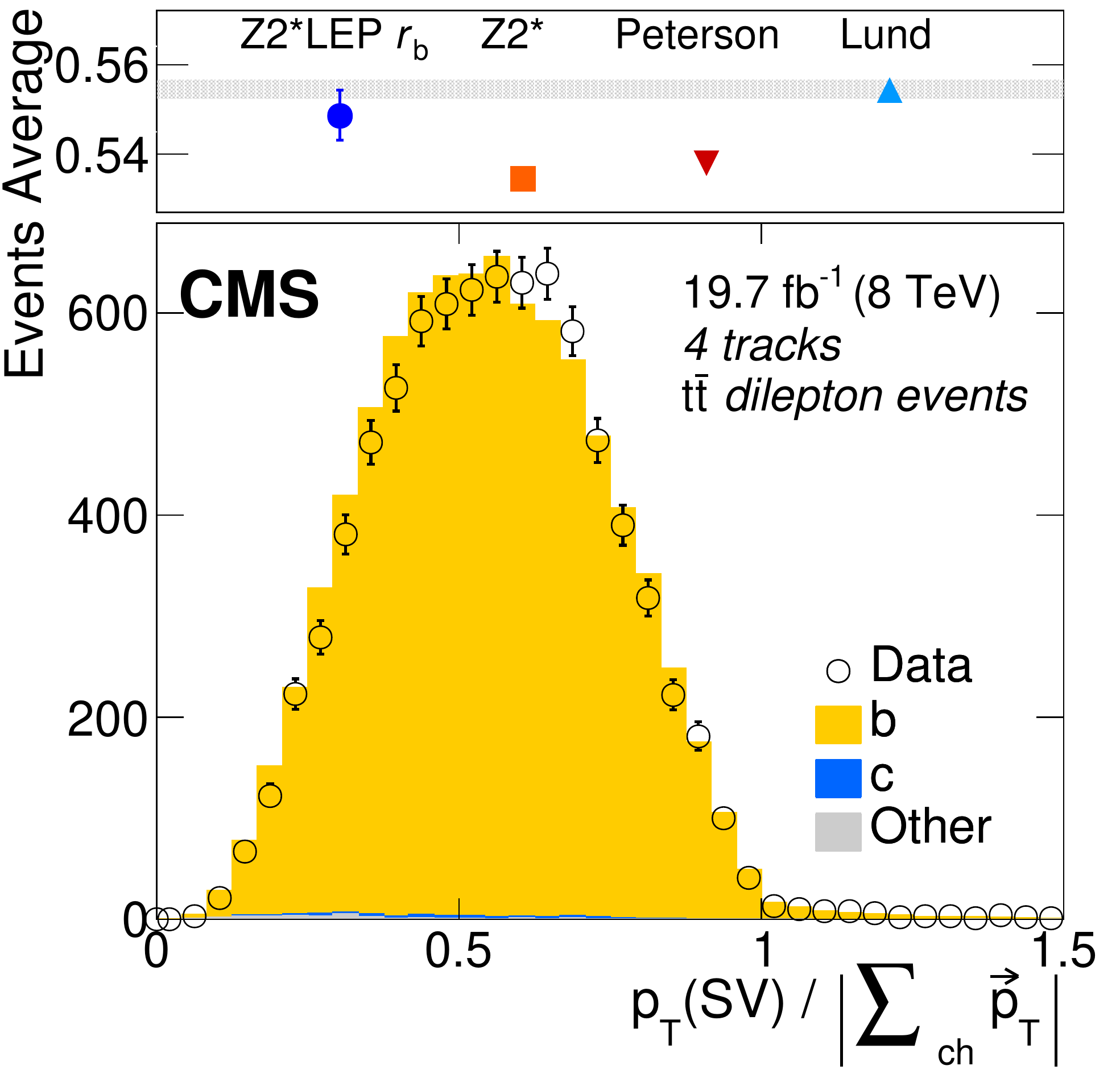}
\includegraphics[width=0.32\textwidth]{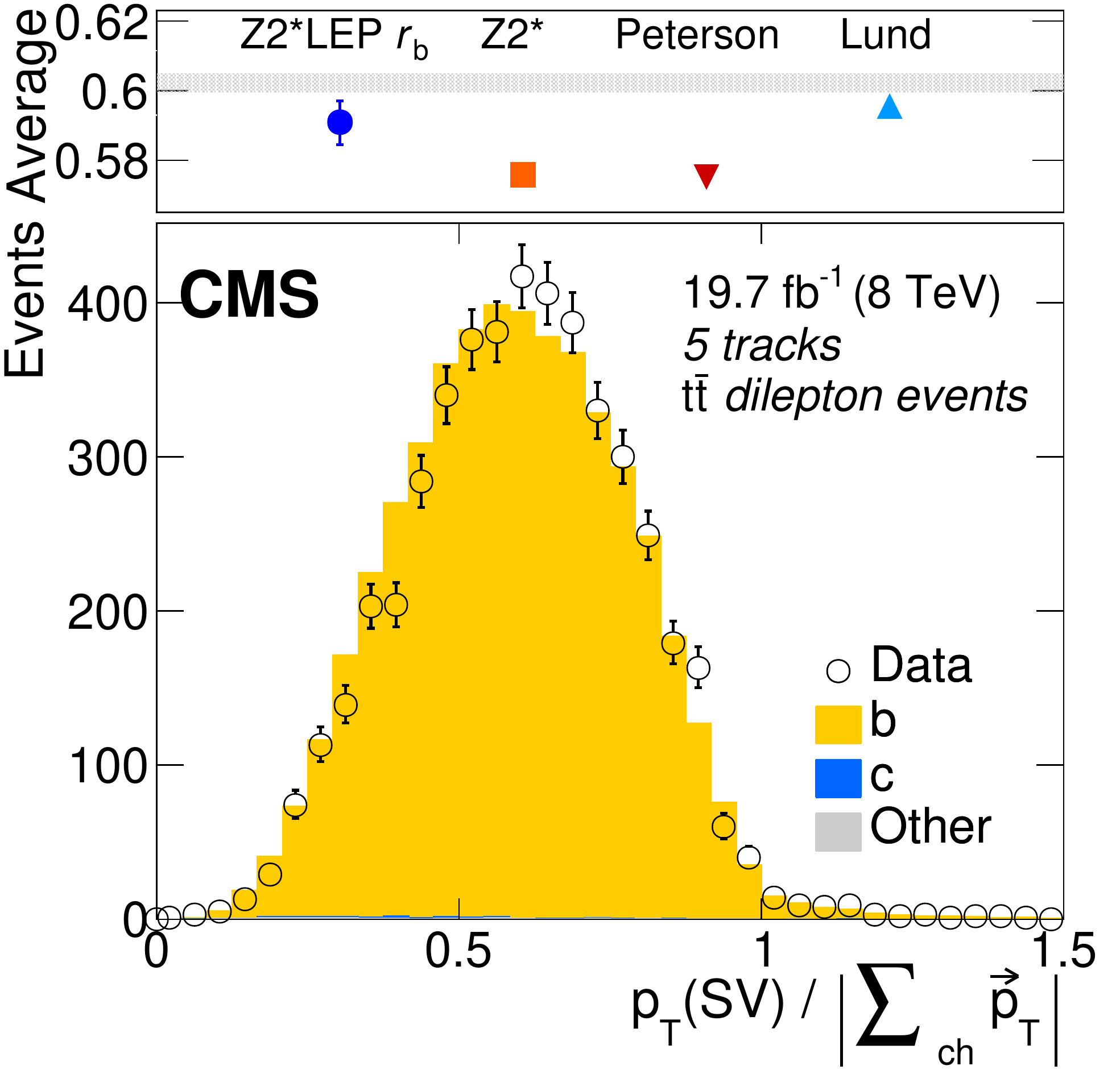}
\caption{
Distributions of the ratio of the transverse momentum of secondary vertices to the charged component of the jet with three, four, and five tracks (from left to right) in $\PZ$+jets dilepton (top) and \ttbar\ dilepton events (bottom), compared to the expected shape using the \ztsrblep\ fragmentation tune. In each plot, the top panels compare the average of the distribution measured in data and its statistical uncertainty (shaded area) with that expected from different choices of the \cPqb\ quark fragmentation function in \PYTHIA.\@ For \ztsrblep, the error bar represents the $\pm$ variations of \ztsrblep.
}
\label{fig:svchfrac}
\end{figure*}

Effects arising from mismodeling of the overall kinematic properties of the event are canceled, to first approximation, by studying the ratio of the two momenta, in which the secondary vertex serves as a proxy for the $\PQb$~hadron and the charged particles represent the full momentum of the initial \cPqb{} quark.
Note that this observable is not sensitive to variations in the jet energy scale, as it makes use only of the charged constituents of the selected jets.
The observed and predicted distributions for $\mathcal{F}_\mathrm{ch}$ in $\PZ$+jets events are shown in Fig.~\ref{fig:svchfrac} (top), separately for vertices with three, four, and five tracks.
For each plot the average of the distribution in the data is compared to the MC prediction using different \cPqb\ fragmentation tunes.
The data appear to favor softer fragmentation shapes such as the \ztwostar\ and Peterson tunes.
However, in this selection a significant fraction of the selected jets stems from the hadronization of light and charm quarks which are not changed by the event reweighting procedure used to compare the different tunes.
Likewise, the \ztsrblep\ tune only affects the simulated fragmentation of \cPqb\ quarks and was obtained using data from LEP enriched in jets from \cPqb\ quark hadronizations, and hence is not expected to correctly describe charm and light quark fragmentation.

In the sample of \ttbar{} events, selected as described in Section~\ref{sec:eventsel}, and used later for the top quark mass extraction, the selected jets are expected to contain a significantly larger fraction of \cPqb{} quarks.
From simulation, we expect a negligible dependence of $\mathcal{F}_\mathrm {ch}$ on the kinematic properties and mass of the top quarks, making this distribution appropriate to compare different fragmentation models.
The equivalent distributions of secondary vertex properties in \ttbar{} events are shown in Fig.~\ref{fig:svchfrac} (bottom).

The observed distributions in this signal selection are generally well described by the central (\ztsrblep) tune, but the comparison of the mean values of $\mathcal{F}_\mathrm {ch}$---as shown in the top panels of the plots---reveals differences between the various fragmentation shapes.
Unlike in the $\PZ$+jets data, the \ztwostar\ tune shows the largest deviation with respect to the \ttbar\ data among the studied variations, whereas the \ztsrblep\ fragmentation shape is in better agreement.
Furthermore, the hard and soft variations of \ztsrblep, corresponding to one standard deviation variations of the $r_{\cPqb}$ parameter, provide a bracketing that encloses or approaches the data.
The \ztsrblep\ tune is therefore used as the nominal \cPqb\ quark fragmentation shape in the following analysis, with the shape variations used to estimate systematic uncertainties in the top quark mass measurement.

\subsection{Inclusive charm mesons \texorpdfstring{in \ttbar{} events}{in t-tbar events}}
Kinematic properties of inclusively reconstructed charmed mesons inside \cPqb\ jets from top quark decays are expected to be sensitive to the modeling of \cPqb\ quark fragmentation.
We limit the study to meson decays with large branching fractions and high expected signal-to-background ratios: $\PJGy{}\rightarrow\Pgmp{}\Pgmm{}$, $\PDz{}\rightarrow\PKm{}\Pgpp{}$ in semileptonic $\PB$ decays, and inclusive $\PDstp{}\rightarrow\PDz{}\Pgpp{}$, with $\PDz{}\rightarrow\PKm{}\Pgpp{}$.

Top quark pair signal events are selected as described above, but with the requirement of at least one \cPqb-tagged jet replacing that of the presence of a reconstructed secondary vertex.
In the dilepton channels the \cPqb\ tagging algorithm output threshold is relaxed, as the expected background is lower.
All five leptonic decay channels of the \ttbar{} state are considered, as discussed above.
To gather as much data as possible, both \cPqb{} jets in each event are considered, selected by their tagging discriminant value and their transverse momentum.
All charged PF candidates used in the jet clustering are used to reconstruct mesons, with particle identification restricted to distinguishing electrons and muons from charged hadrons.

\begin{figure*}[htp]
\centering
\includegraphics[width=0.32\textwidth]{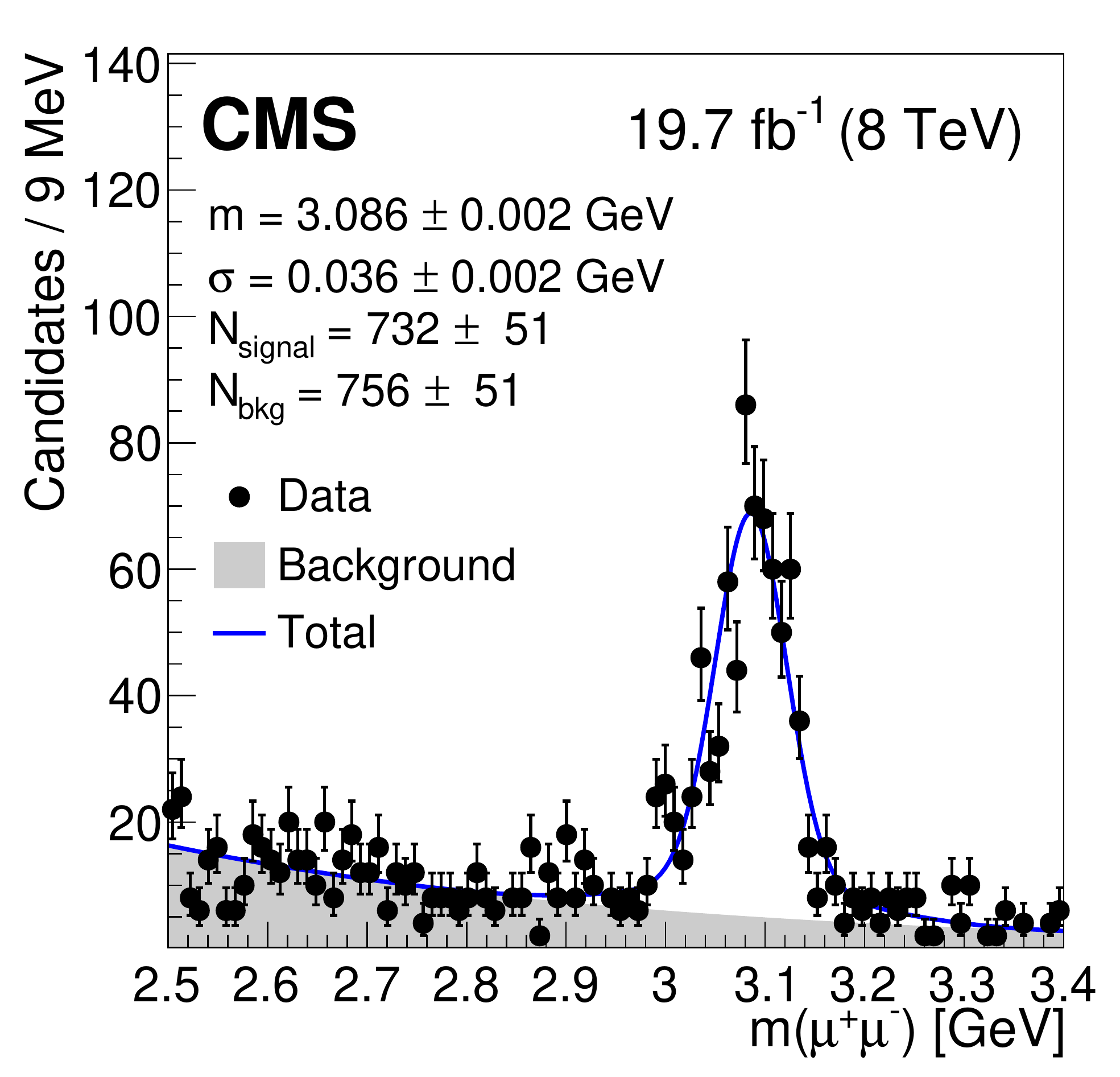}
\includegraphics[width=0.32\textwidth]{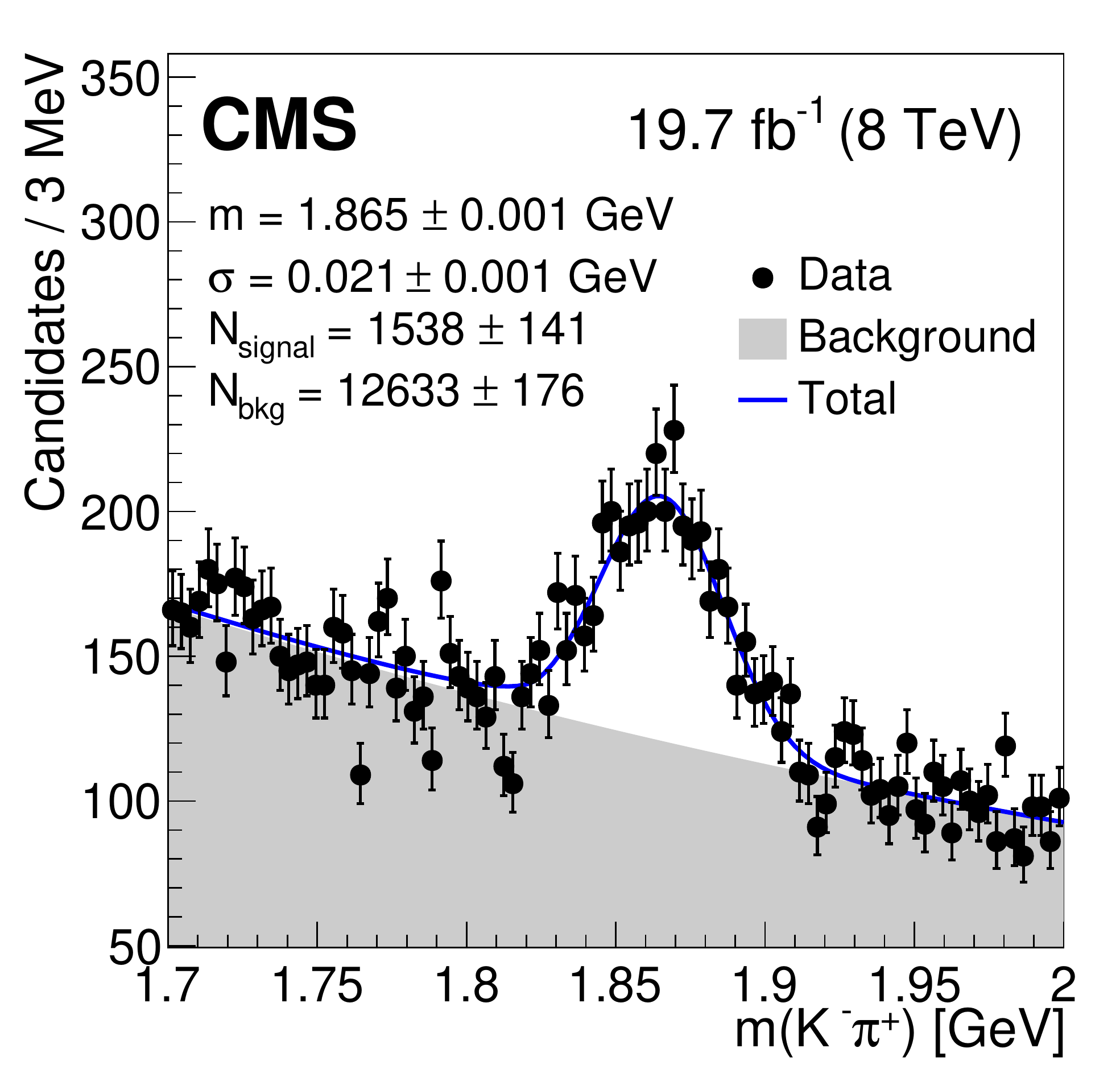}
\includegraphics[width=0.32\textwidth]{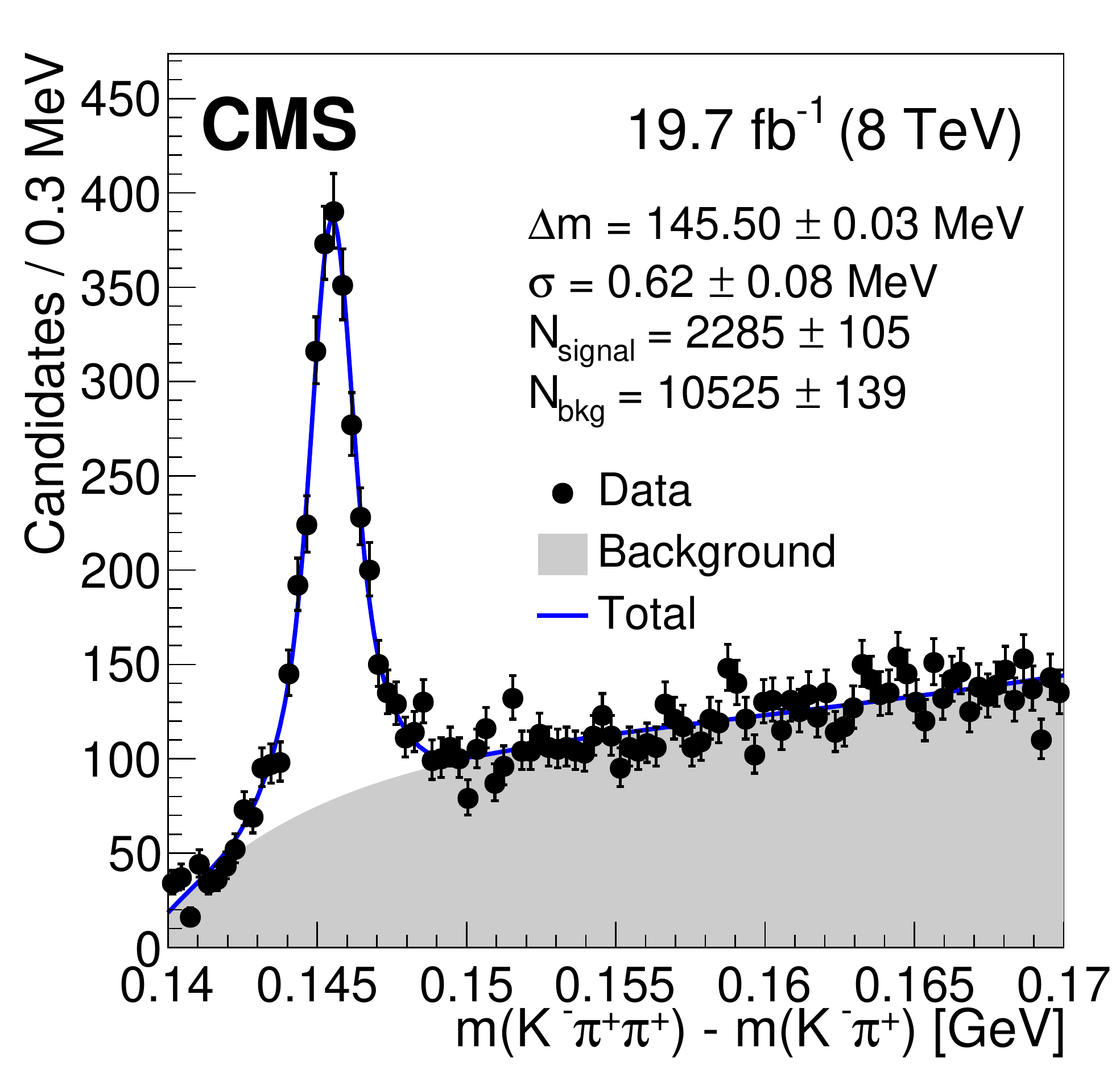}
\caption{
Fits to the invariant mass peaks of the three considered charmed mesons in \ttbar{} events in the data, as described in the text: \PJGy{} (left), \PDz{} (middle), and \PDstp{} (right).
}
\label{fig:charmfits}
\end{figure*}

Candidates for \PJGy{} mesons are reconstructed by requiring two opposite-sign muon candidates among the charged jet constituents, and fitting their invariant mass in the range of 2.5--3.4\GeV, as shown in Fig.~\ref{fig:charmfits}.
The distribution is modeled with the sum of two Gaussian functions for the \PJGy{} signal and a falling exponential for the combinatorial backgrounds.

Neutral charm mesons, \PDz{}, are produced in the majority of $\PQb$~hadron decays, and are reconstructed via their decay to a \PKm{} and \Pgpp{}.
To reduce combinatorial backgrounds they are selected together with a soft lepton from a semileptonic $\PQb$~hadron decay, whose charge determines the respective flavor of the two hadron tracks.
All opposite-sign permutations of the three leading charged constituents of the jet are considered for \PK{} and \Pgp{} candidates and no additional vertex reconstruction is attempted.
The \PK{}\Pgp{} invariant mass is then fitted between 1.7 and 2.0\GeV, using a Crystal Ball~\cite{SLAC-R-236} shape for the signal and an exponential for the combinatorial backgrounds, as shown in Fig.~\ref{fig:charmfits}.

A large fraction of \PDz{} mesons is produced in the decays of intermediate excited charmed hadron states, such as the \PDstp{}, which can be reconstructed by considering the difference in invariant mass between the three-track (\PK{}\Pgp{}\Pgp{}) and the two-track (\PK{}\Pgp{}) systems, where a soft pion is emitted in the $\PDstp\rightarrow\PDz{}\Pgpp{}$ decay.
The \PDz{} mesons are reconstructed among the three leading tracks as described in the previous paragraph, and selected in a mass window of 50\MeV{} around the nominal \PDz{} mass.
A third track of the same charge as the \Pgp{} candidate from the \PDz{} decay is then added, and the mass difference is fitted in a range of 140--170\MeV{}, as shown in Fig.~\ref{fig:charmfits}.
The shape of the mass difference showing the \PDstp{} resonance is modeled using a sum of two Gaussian functions for the signal and a threshold function for the combinatorial backgrounds.

The position of the fitted invariant mass peaks---reconstructed purely in the silicon tracker---agree with the expected meson rest masses within about $0.05\%$ for the \PDz{} and \PDstp{}, indicating that the pion and kaon momentum scales are very well described.
The observed \PJGy{} meson mass, reconstructed using muons, agrees with the expectation~\cite{Agashe:2014kda} within about $0.3\%$, well within the muon momentum scale uncertainty.

\begin{figure*}[htp]
\centering
\includegraphics[width=0.32\textwidth]{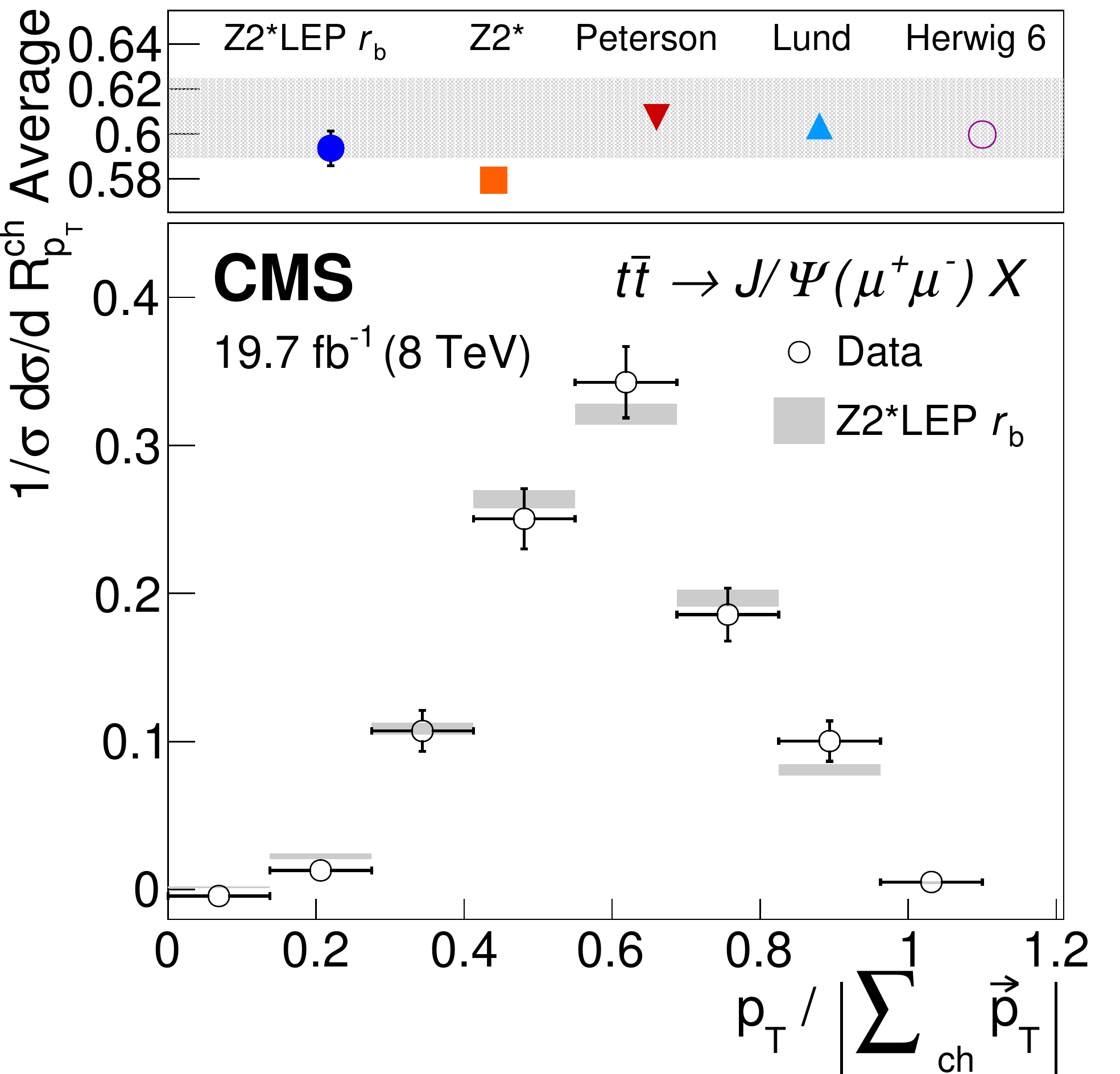}
\includegraphics[width=0.32\textwidth]{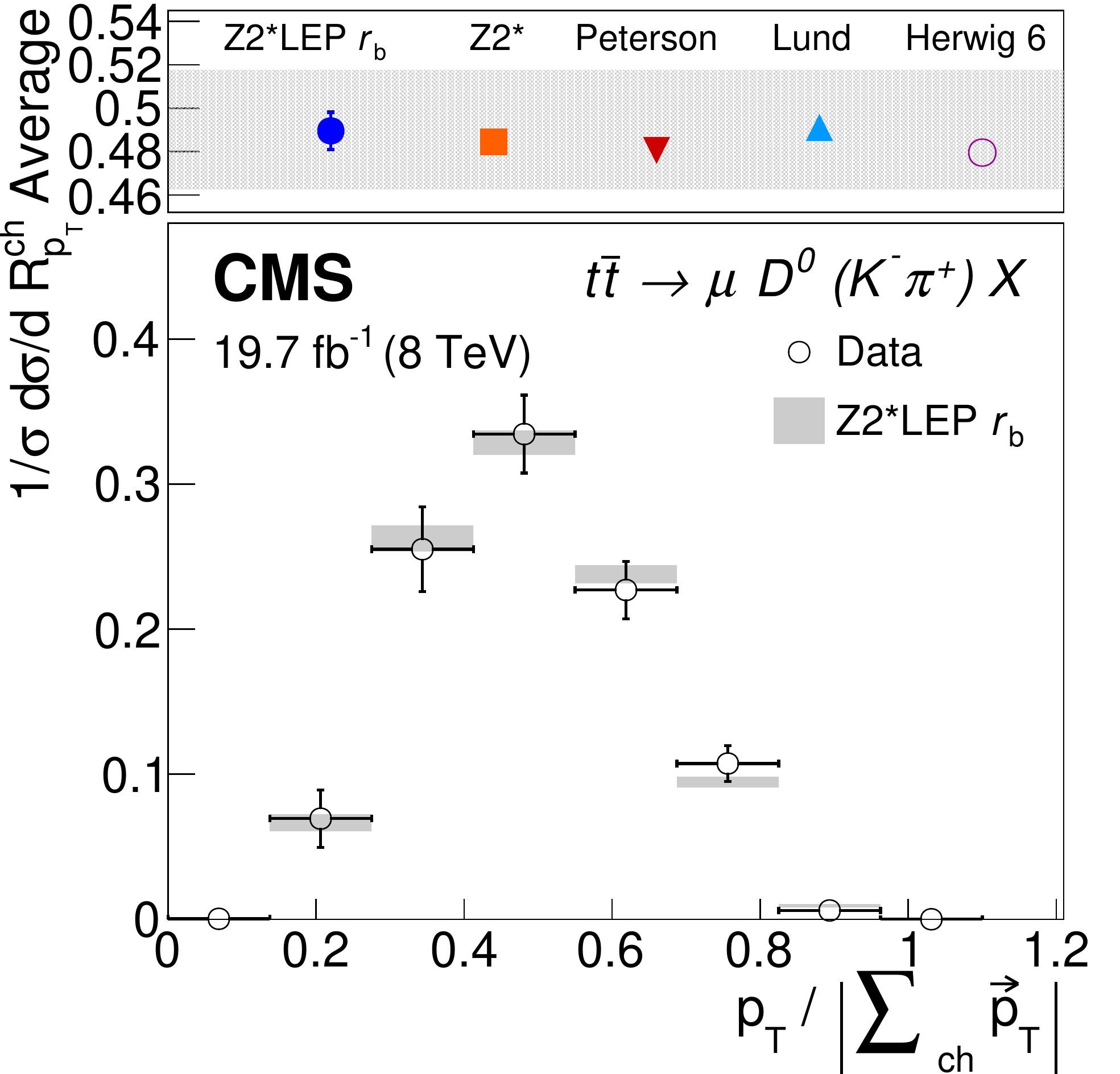}
\includegraphics[width=0.32\textwidth]{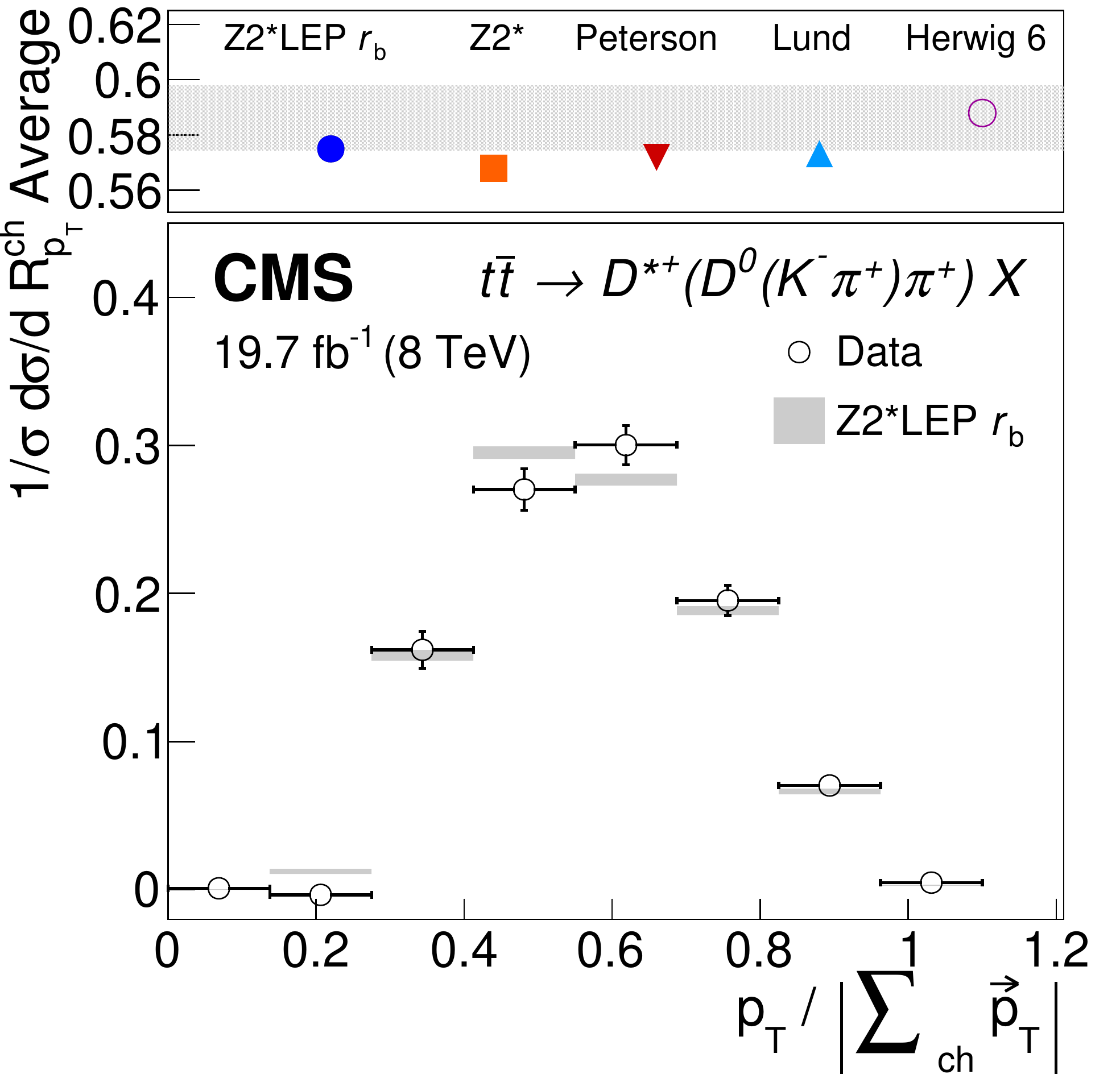}
\caption{
Distribution of the relative transverse momentum of \PJGy{} (left), \PDz{} (middle), and \PDstp{} (right) meson candidates with respect to the charged components of the jet in \ttbar{} events for the data and the nominal \ztsrblep\ fragmentation function.
The top panels show the average of the distributions observed in the data and its statistical uncertainty (shaded area), as well as expectations obtained with different \cPqb\ quark fragmentation functions and with an alternative generator setup using \HERWIG\ 6 with the AUET2 tune.
}
\label{fig:charmchfrac}
\end{figure*}

The fitted signal and background distributions are then used to extract the kinematic properties of the reconstructed mesons using the $_{s}{\mathcal Plot}$ technique~\cite{2005NIMPA.555..356P}, where a discriminating observable (in this case the invariant mass of the candidates) is used to separate the signal and background contributions to the distribution of an observable of interest.
The same method is applied to simulated events with different generator tunes and a range of different \cPqb\ quark fragmentation functions, and the results are compared
with data.
Among several investigated kinematic properties of the charm meson candidates, the fraction of transverse momentum relative to the charged component of the jet momentum shows the highest sensitivity to variations in the \cPqb\ quark fragmentation shape.
The results are displayed in Fig.~\ref{fig:charmchfrac}.

The reconstructed mesons are observed to carry about 50--60\% of the overall charged jet momentum.
These results are in good agreement with the predictions obtained from simulated \ttbar{} events for the central fragmentation function choice and corresponding variations.
The conclusions from the study of secondary vertex properties in the previous section are confirmed by the charm meson properties, with the \ztsrblep\ fragmentation showing better agreement with the data than the nominal \ztwostar\ shape, albeit with a large statistical uncertainty.

The numbers of meson candidates observed in the data are reproduced within about 10\% when \PYTHIA\ with the \ztwostar\ tune is used in the parton shower and hadronization, whereas \HERWIG~6~\cite{Corcella:2000bw} with the AUET2 tune~\cite{ATL-PHYS-PUB-2011-008} underestimates both the \PDstp{} and \PJGy{} yields by more than 50\%, and overestimates \PDz{} production by about 30\%.

\section{Top quark mass measurement}\label{sec:topmass}
Observables that are dependent on the top quark mass are constructed using the kinematic properties of the decay products of the top quark.
The choice of observable is a compromise between sensitivity to the mass on the one hand and susceptibility to systematic uncertainties on the other hand.
The most precise measurements to date have approached this trade-off by fully reconstructing the top quark from three jets in hadronic decays, heavily relying on precise calibrations of the reconstructed jet energies.
In the analysis presented here, a different approach is used that sacrifices some sensitivity to minimize the reliance on detector calibrations.
This exposes the result to uncertainties in the modeling of top quark decays and \cPqb{} hadronization, but has reduced experimental uncertainties.
The analysis will therefore immediately benefit from a future improvement of our understanding of these effects.

\subsection{Observable and measurement strategy}\label{ssec:strategy}
The observable exploited in this analysis is built from the measured properties of the charged lepton from the \PW~boson decay and the charged constituents of a hadronic jet compatible with originating from a common secondary vertex.
The invariant mass of the secondary vertex-lepton system, \msvl, then serves as a proxy for the top quark mass.
In building the invariant mass, the vertex constituents are assumed to be charged pions.
The \msvl\ variable shows a strong dependence on the mass of the top quark despite not accounting for the neutrino from the \PW~boson decay or from semileptonic $\PQb$~hadron decays, nor for neutral products of the \cPqb~quark hadronization.
Using only charged particles and well-modeled leptons reduces the main experimental uncertainties to acceptance effects.

For each selected event, all possible combinations of leptons and secondary vertices---up to two in semileptonic events and up to four in dileptonic events---are taken into account in the measurement.
Hence, by construction, the same number of correct and wrong combinations (\ie\ pairing the lepton with the vertex associated with the other top quark decay) enter the analysis.
In simulation, in about 11\% of cases the selected vertex could not be attributed to the decay products of either \cPqb\ quarks and is most likely spurious, either from a light quark from a hadronic \PW~boson decay, or from a gluon or light quark from initial-state radiation.

\begin{figure}[htp]
\centering
\includegraphics[width=0.45\textwidth]{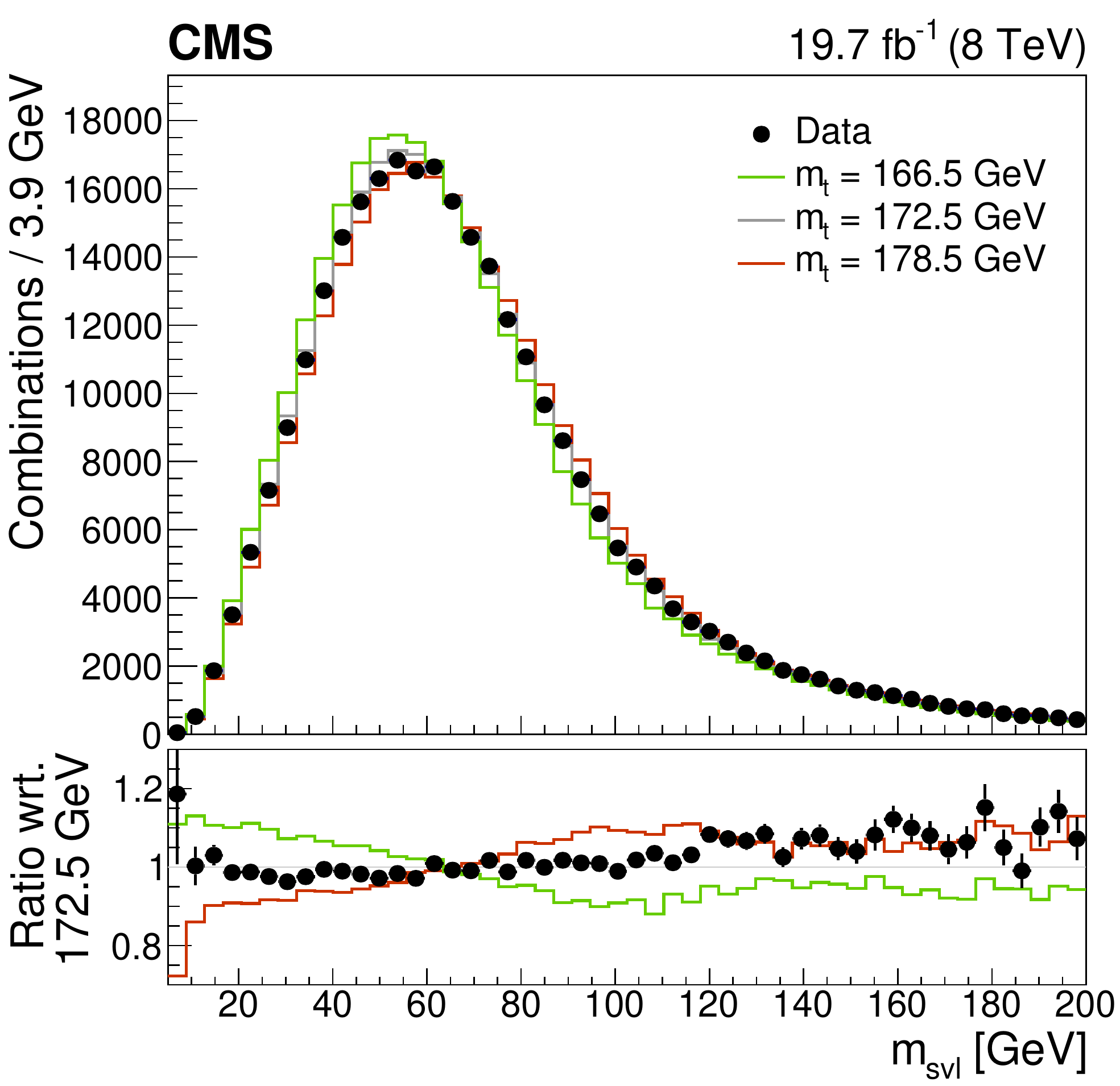}
\caption{
Lepton-SV invariant mass distribution for a combination of all five channels, for a simulation of three different top quark mass values (166.5, 172.5, and 178.5\GeV), and the observed data distribution.
Note that all possible lepton-vertex combinations for each event enter the distribution.
}
\label{fig:msvldatamc}
\end{figure}

Figure~\ref{fig:msvldatamc} shows the observed \msvl\ distribution for a combination of all five channels, compared to simulated distributions at three different generated top quark mass values.

The shape of the \msvl{} observable depends considerably on the number of tracks associated with the secondary vertex, shifting to higher values as more tracks are included.
The analysis is therefore carried out in three exclusive track multiplicity categories of exactly three, four, or five tracks.
Vertices with only two tracks show an increased level of backgrounds and reduced sensitivity to \mtop{} and are therefore excluded from the analysis.
Furthermore, when evaluating systematic uncertainties, the results from the individual categories are assigned weights corresponding to the observed event yields in each, to absorb any mismodeling of the vertex multiplicity distribution in simulated events.
Hence the analysis is carried out in fifteen mutually exclusive categories---three track multiplicities and five lepton flavor channels---and combined to yield the final result.

\subsection{Signal and background modeling}\label{ssec:signalmodel}
The observed \msvl{} distributions in each category are fitted with a combination of six individual components:
\begin{itemize}
	\item[--] ``correct'' pairings for the \ttbar{} signal where leptons and vertices are matched to the same top quark decay;
	\item[--] ``wrong'' pairings for the \ttbar{} signal where leptons and vertices are matched to the opposite top quark decay products;
	\item[--] ``unmatched'' pairings for the \ttbar{} signal where leptons are paired with vertices that cannot be matched to a \cPqb{} quark hadronization, \ie{} either from a hadronic \PW~boson decay or from initial- or final-state radiation;
	\item[--] ``correct'' pairings for the single top quark signal;
	\item[--] ``unmatched'' pairings for the single top quark signal, where there can be no ``wrong'' pairs in the sense of the above;
	\item[--] leptons and vertices from background processes.
\end{itemize}
Among those, the ``correct'' pairings both for \ttbar{} and single top quarks, and the ``wrong'' pairings in the \ttbar{} signal carry information about the top quark mass and are parametrized as a function of \mtop.
The relative fractions of correct, wrong, and unmatched pairings for both \ttbar{} and single top quarks and their dependence on \mtop{} are determined from simulated events.
Furthermore, the relative contributions of \ttbar{} and single top quark events are calculated using the top quark mass-dependent theoretical predictions of the production cross sections at NNLO for \ttbar{}, and single top quark $t$ channel as well as $ \PQt \PW $ channel.
The overall combined signal strength of \ttbar{} and single top quark signal is left floating in the final fit, together with \mtop{}.

The background contribution is a combination of different processes, depending on the channel, with dominant contributions from DY+jets in the dilepton channels, and $\PW$+jets and QCD multijet processes in the semileptonic channels.
The overall background yields are fixed to the predictions from simulation, with the exception of QCD multijets, the normalization of which is determined from a fit to the \MET{} distribution in the data, and DY+jets, which is normalized in a data control sample selecting dilepton pairs compatible with a \cPZ~boson decay.
The total (statistical plus systematic) uncertainty in the normalization of the QCD multijets and DY+jets backgrounds is about 30\%.

For each channel and track multiplicity category, the full signal model is given by:
\begin{equation*}
 \begin{split}
 \ifthenelse{\boolean{cms@external}}
 {
N & \big[\msvl|\mtop,\mu,\theta_{\rm bkg}\big]\:=\: \\
&\mu N_{\rm top}^{\rm exp} \Big[\alpha^{\rm cor} f_{\ttbar}^{\rm cor}(\msvl|\mtop) + \alpha^{\rm wro} f_{\ttbar}^{\rm wro}(\msvl|\mtop)\\
& \qquad \qquad +(1-\alpha^{\rm cor}-\alpha^{\rm wro})f_{\ttbar}^{\rm unm}(\msvl) \\
& + \kappa_{\cPqt{}}\big[ \alpha^{\rm cor}_{\cPqt{}}  f_{\cPqt{}}^{\rm cor}(\msvl|\mtop) + (1-\alpha^{\rm cor}_{\cPqt{}})  f_{\cPqt{}}^{\rm noncor}(\msvl)\big] \Big] \\
+ & \: N_{\rm bkg}^{\rm exp}(1+\theta_{\rm bkg}) f_{\rm bkg}(\msvl), \\
}{
N\big[\msvl|\mtop,\mu,\theta_{\rm bkg}\big]\:=\:
&\mu N_{\rm top}^{\rm exp} \Big[\alpha^{\rm cor} f_{\ttbar}^{\rm cor}(\msvl|\mtop)+\alpha^{\rm wro} f_{\ttbar}^{\rm wro}(\msvl|\mtop)\\
&\qquad\qquad\;+(1-\alpha^{\rm cor}-\alpha^{\rm wro})f_{\ttbar}^{\rm unm}(\msvl) \\
&\qquad\qquad\;+\kappa_{\cPqt{}}\big[ \alpha^{\rm cor}_{\cPqt{}}  f_{\cPqt{}}^{\rm cor}(\msvl|\mtop) + (1-\alpha^{\rm cor}_{\cPqt{}})  f_{\cPqt{}}^{\rm noncor}(\msvl)\big] \Big] \\
&+N_{\rm bkg}^{\rm exp}(1+\theta_{\rm bkg}) f_{\rm bkg}(\msvl), \\
}
\end{split}
\end{equation*}\
where $N_{\rm top}^{\rm exp}$ and $N_{\rm bkg}^{\rm exp}$ are the number of top quark events (\ttbar{} and single top quarks) and background events expected from simulation;
the $f_{k}^{i}$
are the six \msvl{} templates of which three are parametrized in \mtop{};
$\alpha^{\rm cor}$, $\alpha^{\rm wro}$, and $\alpha^{\rm cor}_{\cPqt}$, are the fractions of correct and wrong lepton-vertex pairings for \ttbar{} and single top quark production, determined from simulated events as a function of \mtop{};
$\kappa_{\cPqt{}}$ is the relative fraction of single top quark events, fixed as a function of \mtop{} from the theoretical prediction;
$\theta_{\rm bkg}$ is a Gaussian penalty for a correction of the background yield;
and finally $\mu$ is the overall signal strength of top quark events, determined in the fit.

The parameters of each of the $f_{k}^{i}$ templates and their possible \mtop\ dependence is determined in a fit to \msvl{} distributions of simulated events in the corresponding category and pairing classification.
The combined background template is built from fits to dedicated samples of simulated events of the corresponding processes, weighted by the expected event yields.
The shape for QCD multijet processes is determined from a control sample of nonisolated leptons in the data and normalized using a fit to the \MET{} distribution.
For correct and wrong pairings in \ttbar{} and for correct pairings in single top quark events, the fit is done for a range of generated top quark mass points in the range 163.5--181.5\GeV, from which a linear dependence of the parameters on \mtop{} is extracted.
The \msvl{} distributions for unmatched pairings and background events do not depend on \mtop.
Each distribution is fitted with the sum of an asymmetric Gaussian ($\mathcal{G}_{\rm asym}$) and a Gamma distribution ($\Gamma$), of which four of the six total parameters are found to provide sensitivity to the top quark mass:
\begin{equation*}
\ifthenelse{\boolean{cms@external}}
{
\begin{split}
f_{k}^{i}(\msvl|\mtop) \;=\;  & \lambda \,  \mathcal{G}_{\rm asym}   \big(\msvl|\mu(\mtop),\sigma_{\rm L}(\mtop),\sigma_{\rm R}(\mtop)\big) \\
&  \:+\: (1-\lambda) \, \Gamma\big(\msvl|\gamma,\beta,\nu(\mtop)\big).
\end{split}
}
{
f_{k}^{i}(\msvl|\mtop) \;=\; \lambda \, \mathcal{G}_{\rm asym}\big(\msvl|\mu(\mtop),\sigma_{\rm L}(\mtop),\sigma_{\rm R}(\mtop)\big) \:+\: (1-\lambda)\, \Gamma\big(\msvl|\gamma,\beta,\nu(\mtop)\big).
}
\end{equation*}

The shape parameters are the mean of the Gaussian peak ($\mu$), the left and right width parameters of the Gaussian ($\sigma_{\rm L}$ and $\sigma_{\rm R}$), the shape parameter of the Gamma distribution ($\gamma$), its scale ($\beta$), and its shift ($\nu$).
Of these, all but $\gamma$ and $\beta$ show some usable sensitivity to the top quark mass.

\begin{figure*}[htp]
\centering
\includegraphics[width=0.32\textwidth]{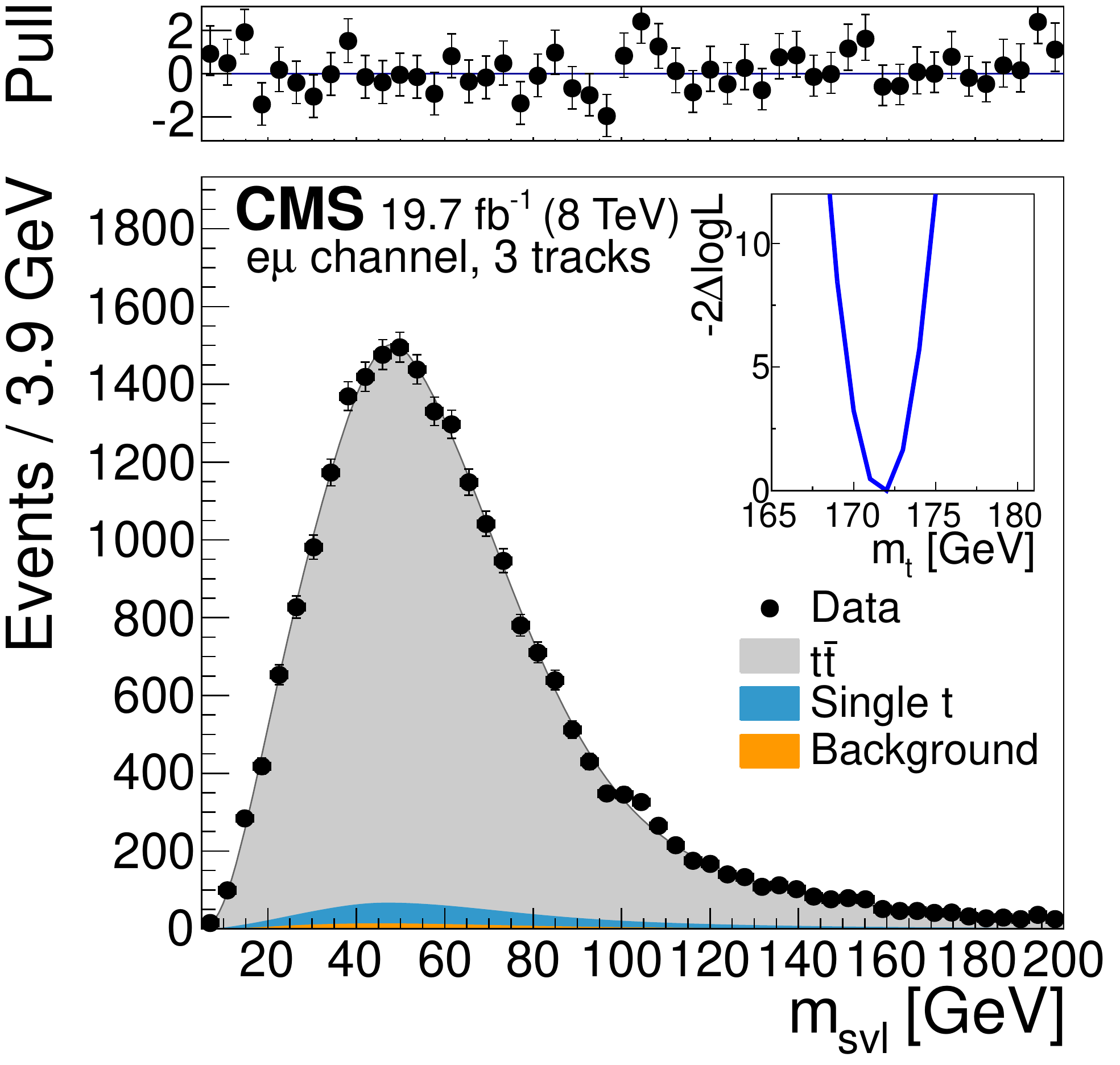}
\includegraphics[width=0.32\textwidth]{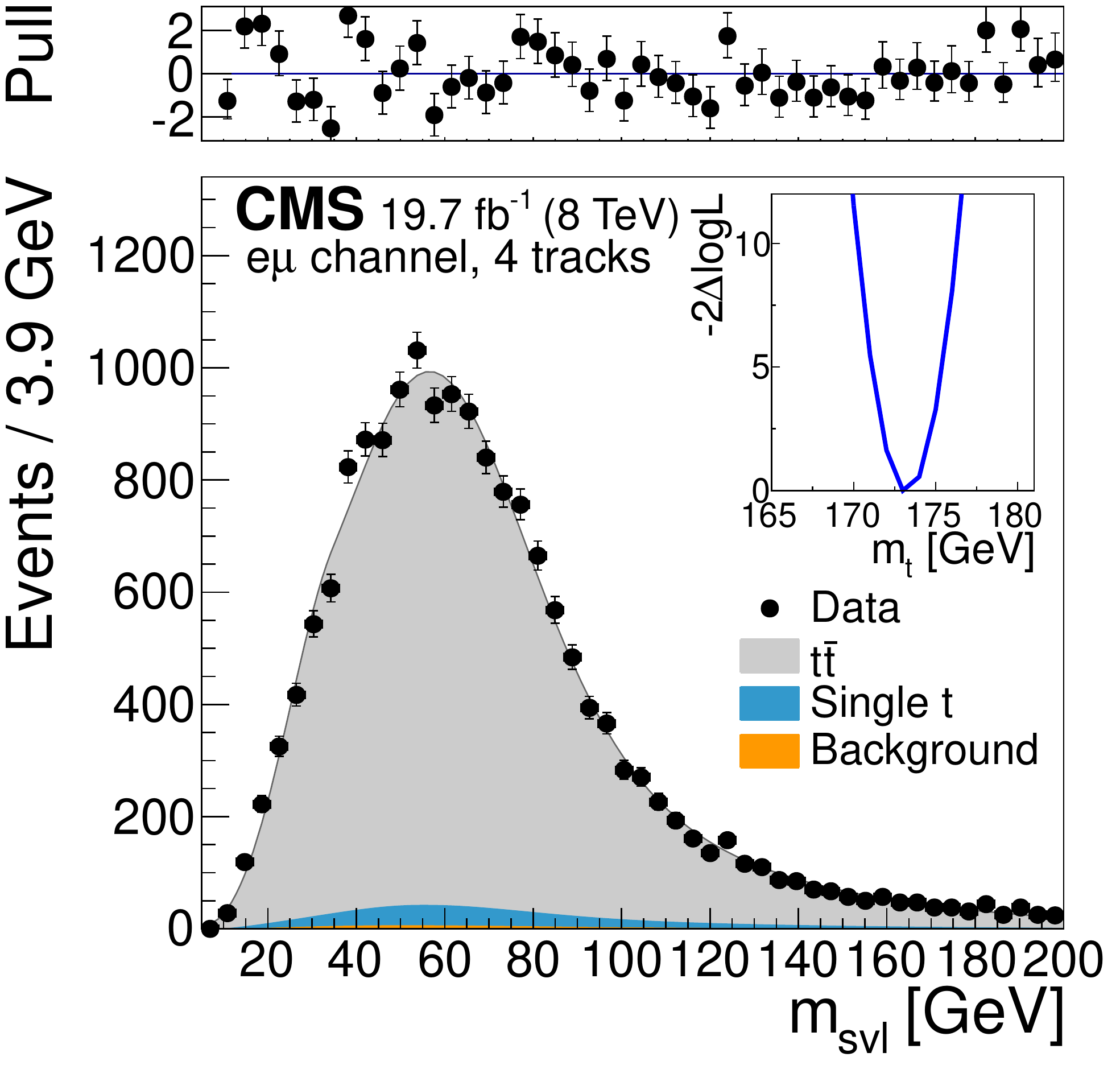}
\includegraphics[width=0.32\textwidth]{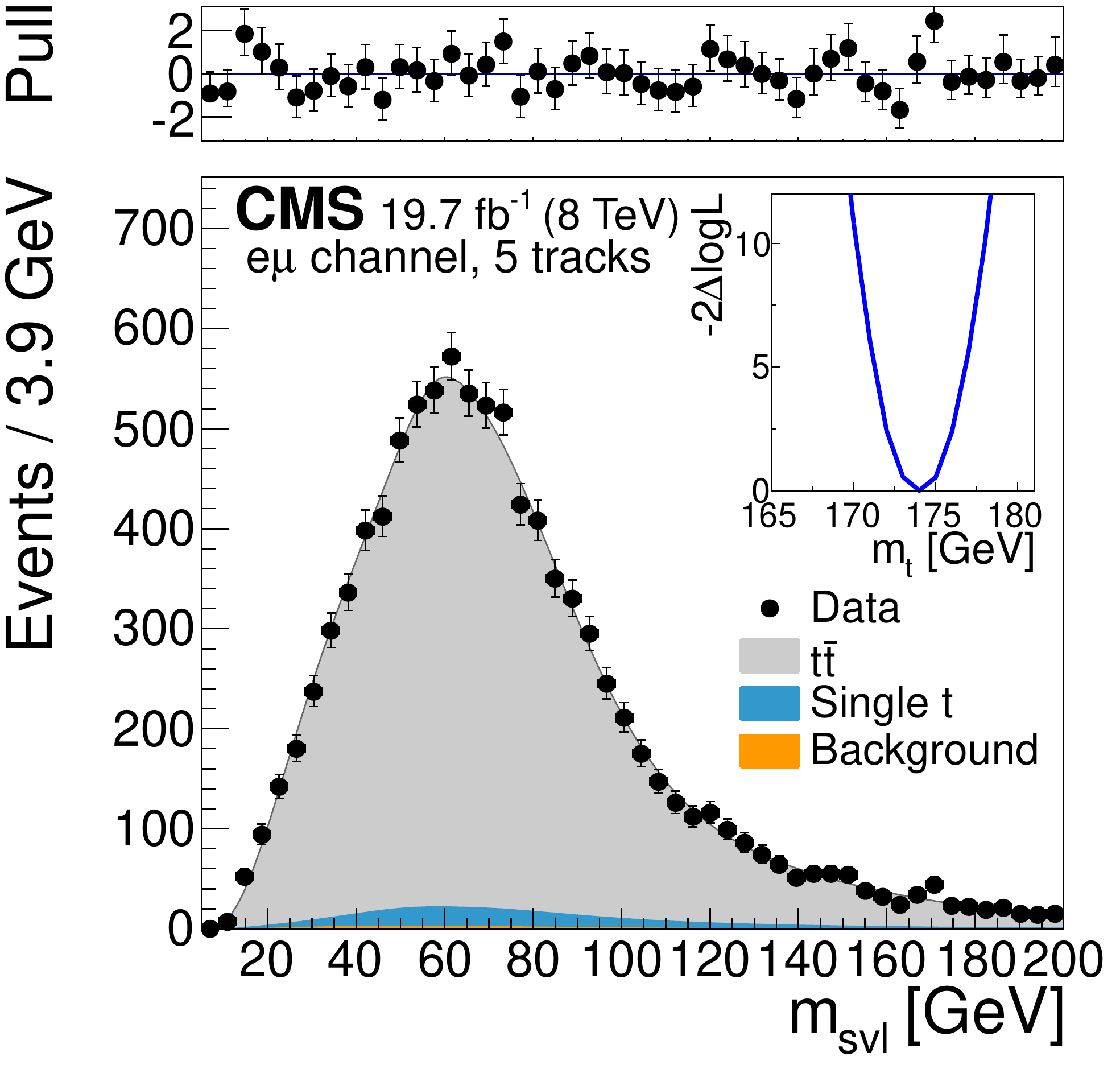}\\
\includegraphics[width=0.32\textwidth]{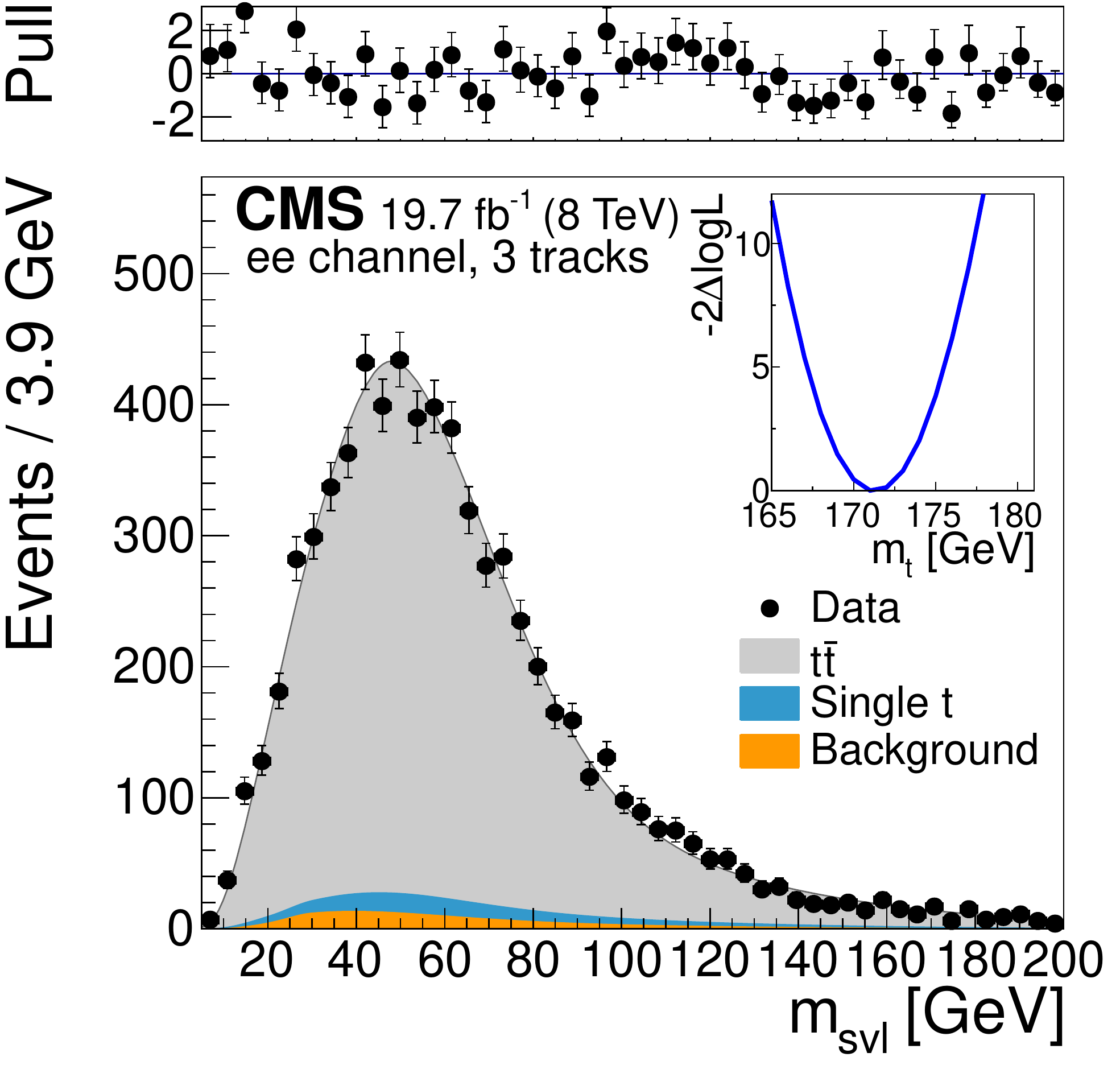}
\includegraphics[width=0.32\textwidth]{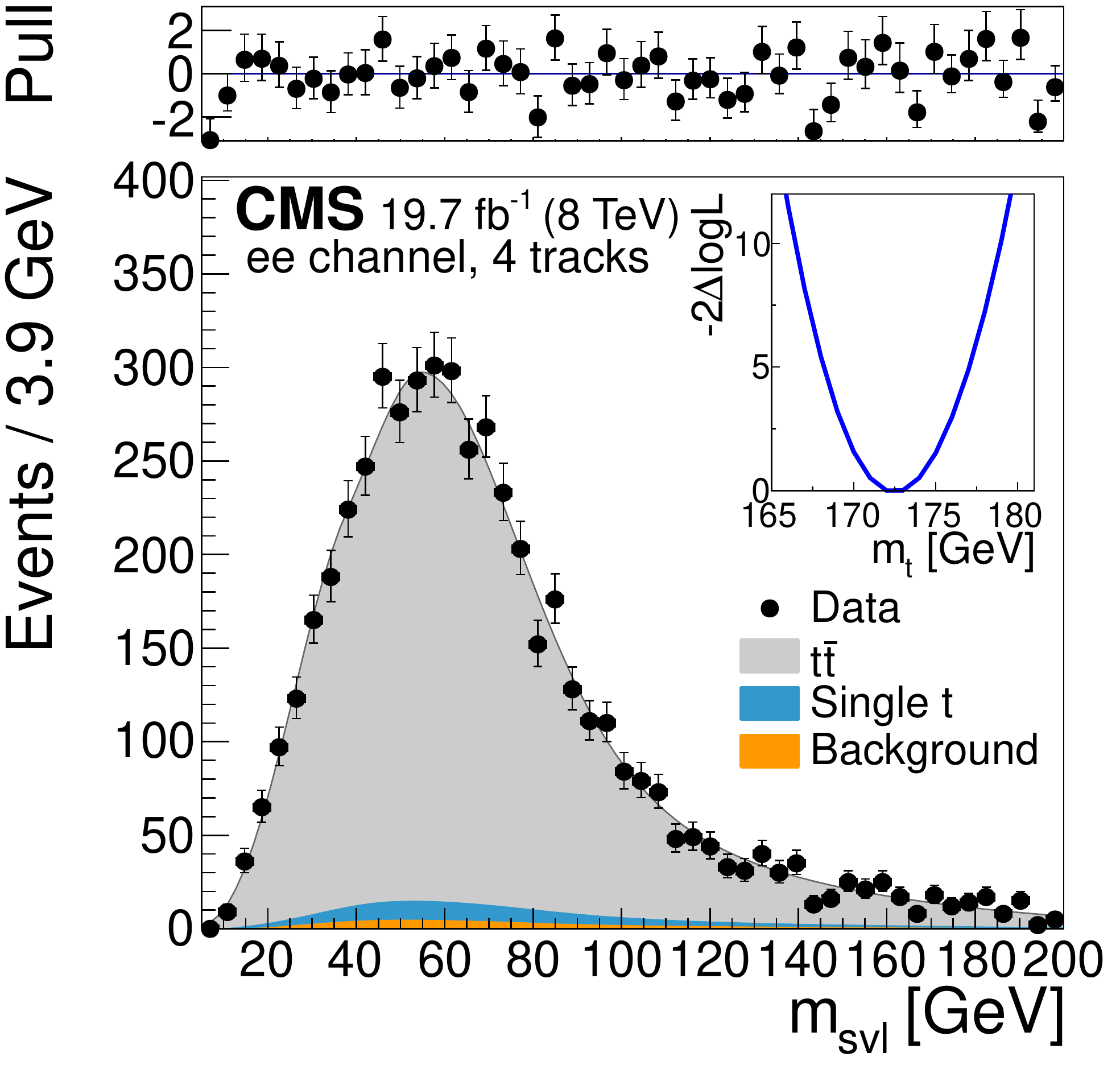}
\includegraphics[width=0.32\textwidth]{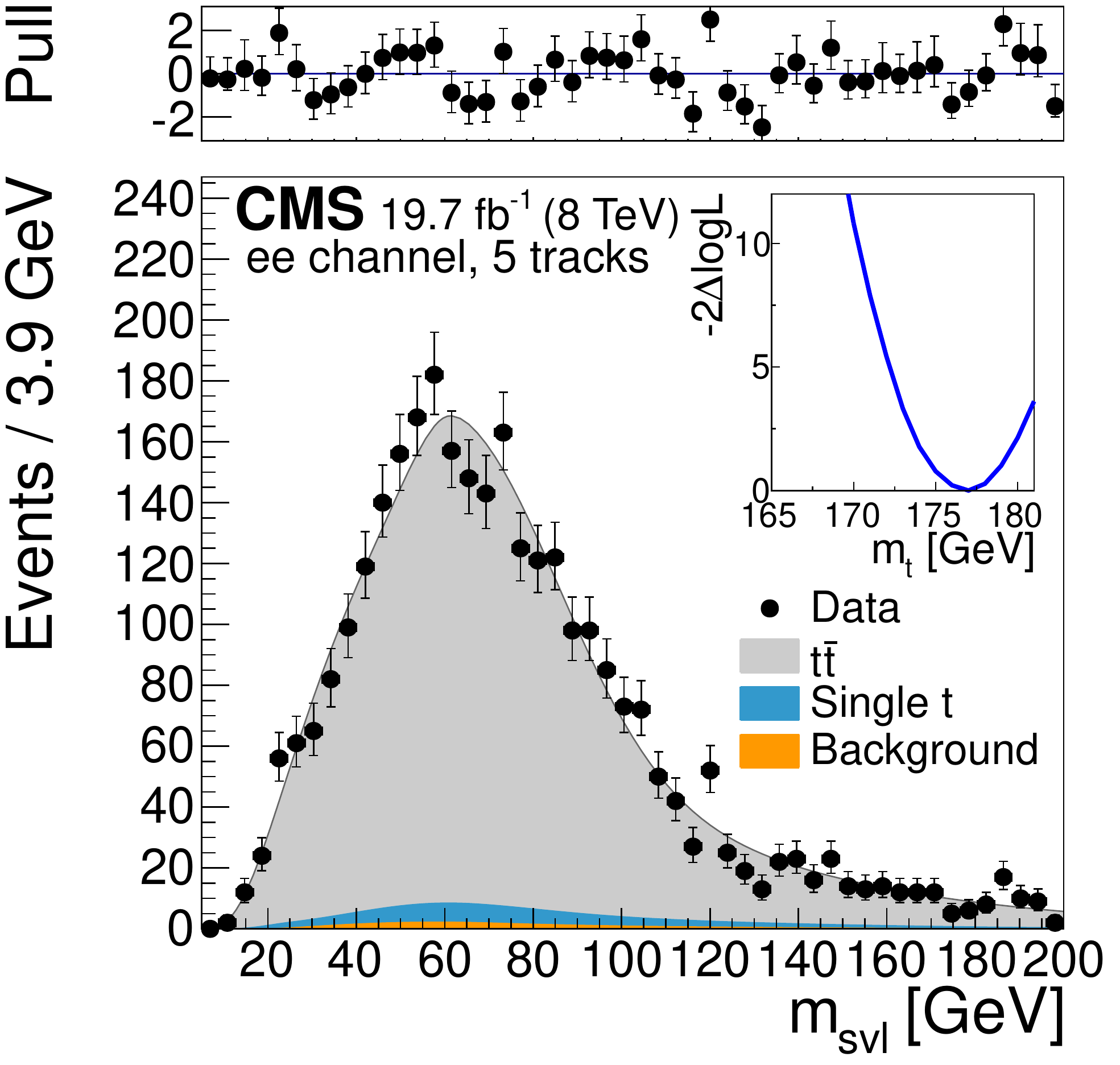}\\
\includegraphics[width=0.32\textwidth]{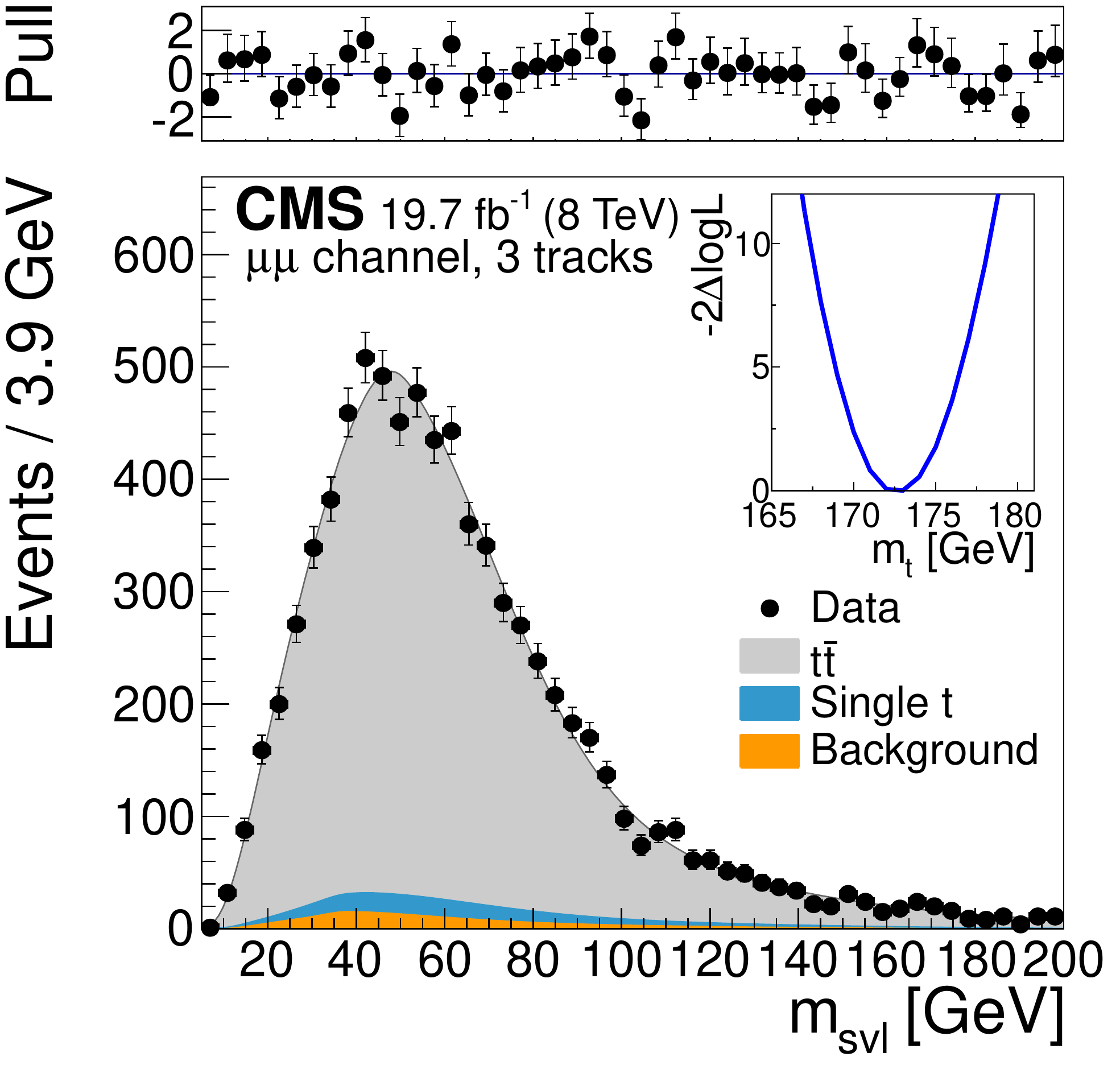}
\includegraphics[width=0.32\textwidth]{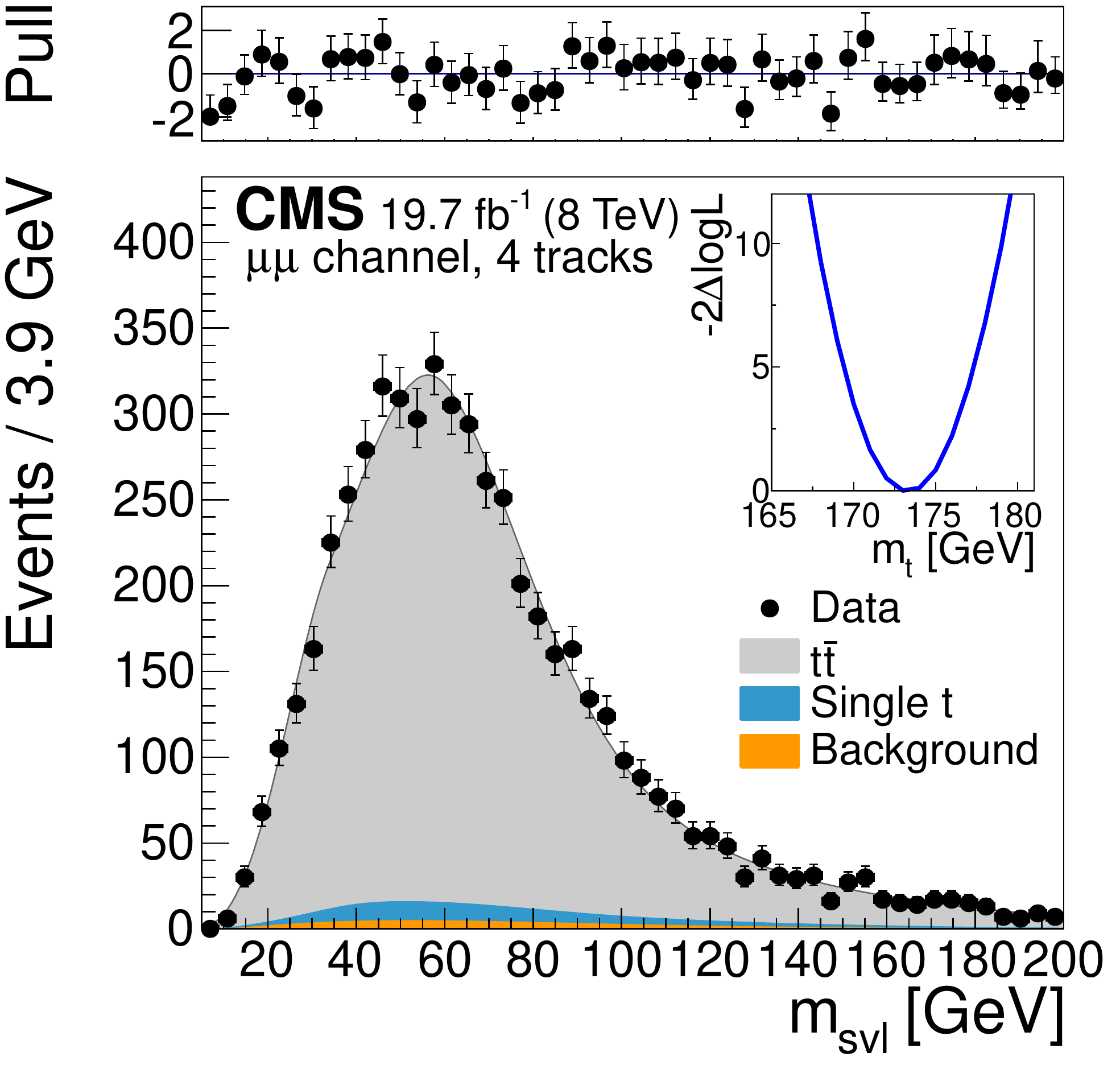}
\includegraphics[width=0.32\textwidth]{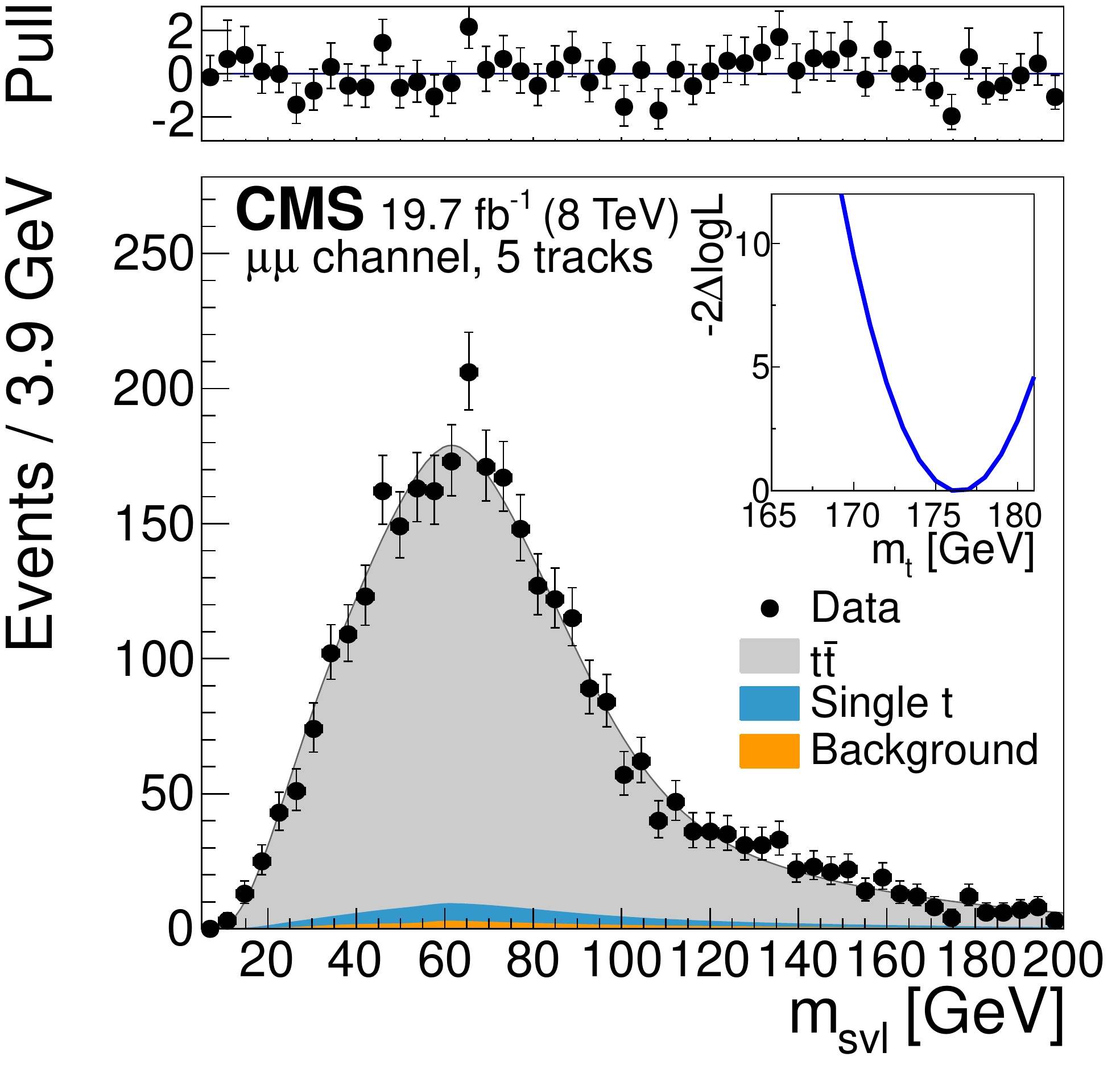}
\caption{
Template fits to the observed \msvl{} distributions for the three dilepton channels (\emu, \ee, \mumu\ from top to bottom row), and for exactly three, four, and five tracks assigned to the secondary vertex (from left to right column).
The top panels show the bin-by-bin difference between the observed data and the fit result, divided by the statistical uncertainty (pull).
The inset shows the scan of the negative log-likelihood as a function of the calibrated top quark mass, accounting only for the statistical uncertainty, when performed exclusively in each event category.
}
\label{fig:msvlfitsdil}
\end{figure*}

\begin{figure*}[htp]
\centering
\includegraphics[width=0.32\textwidth]{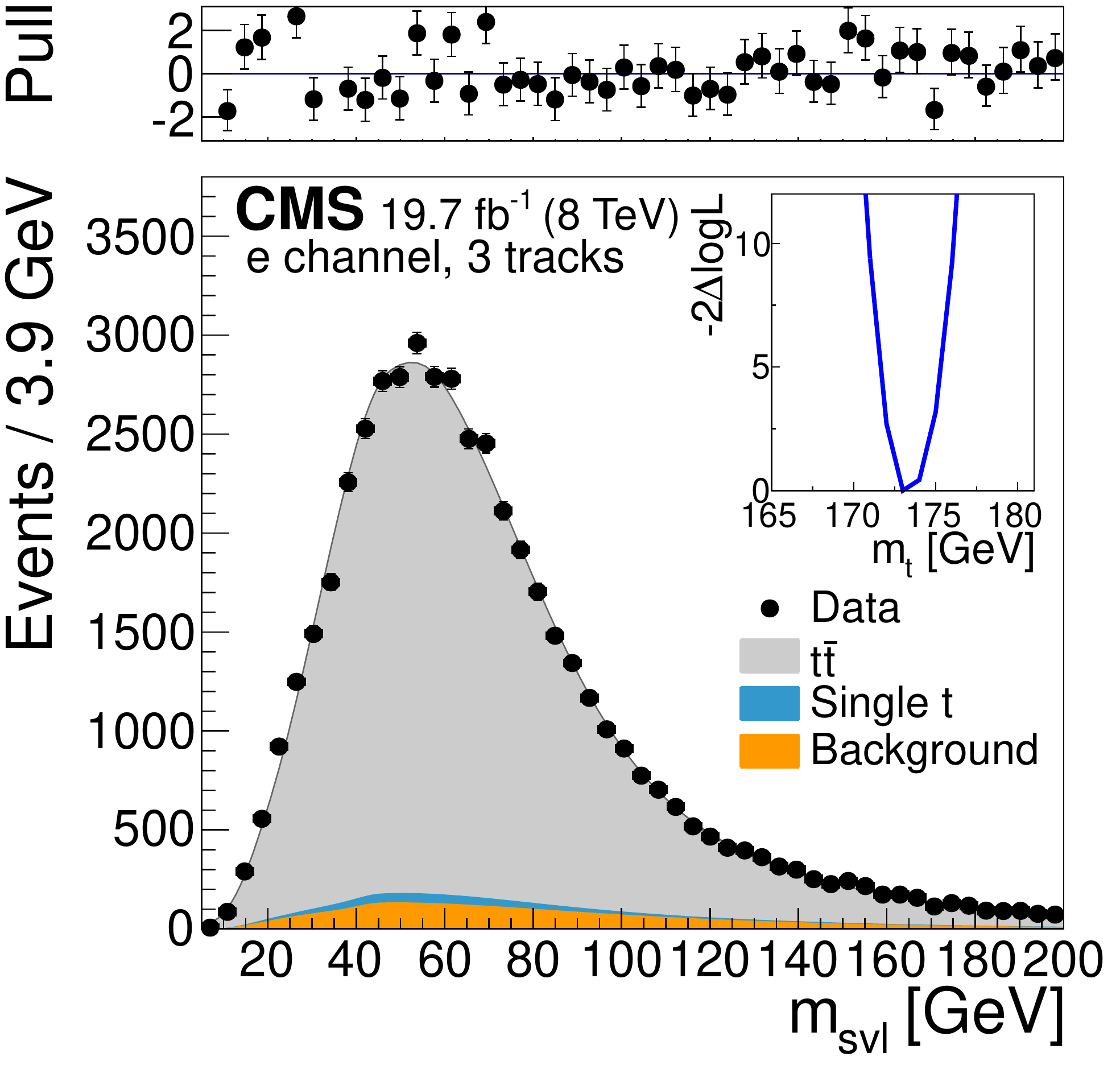}
\includegraphics[width=0.32\textwidth]{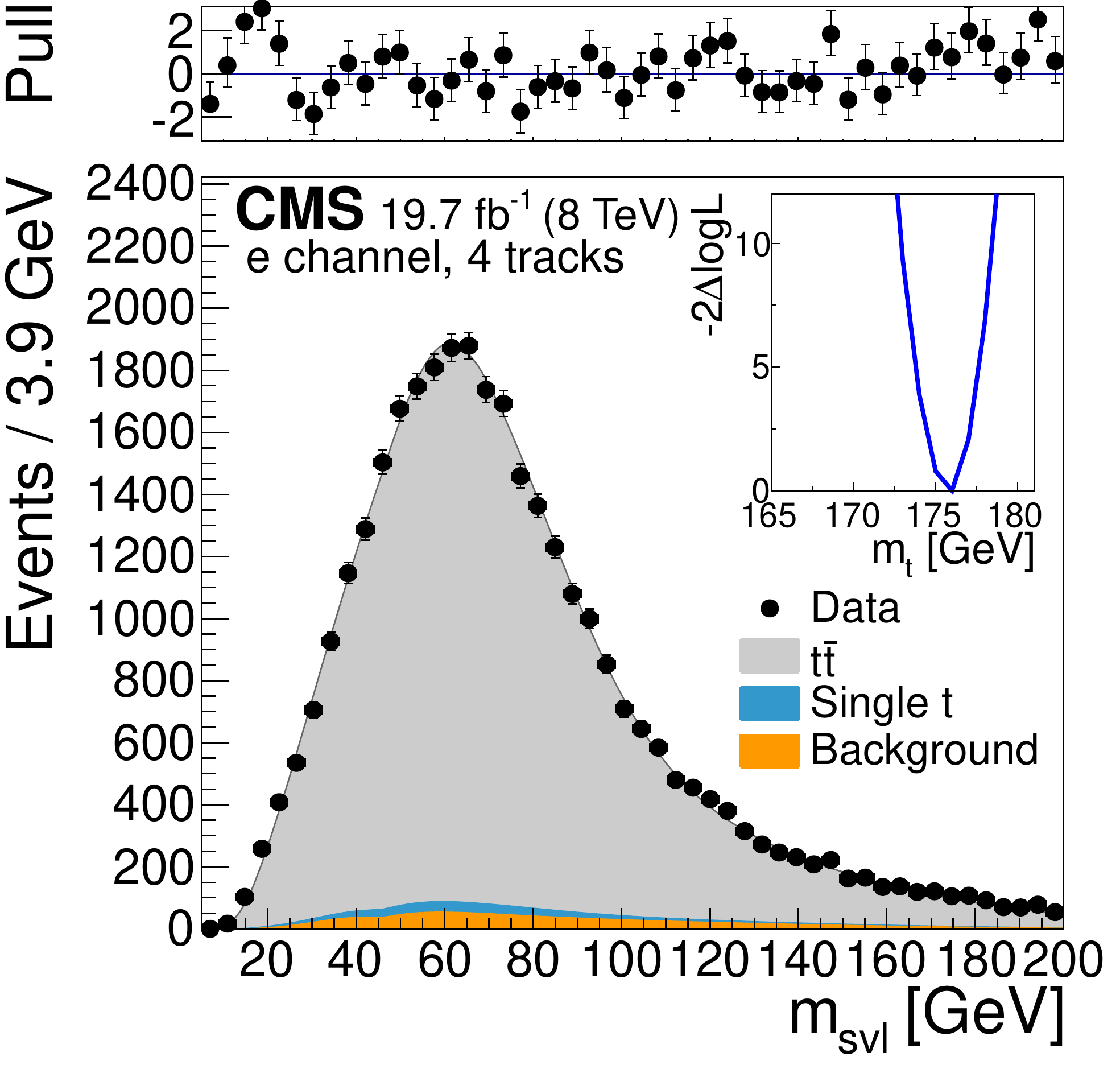}
\includegraphics[width=0.32\textwidth]{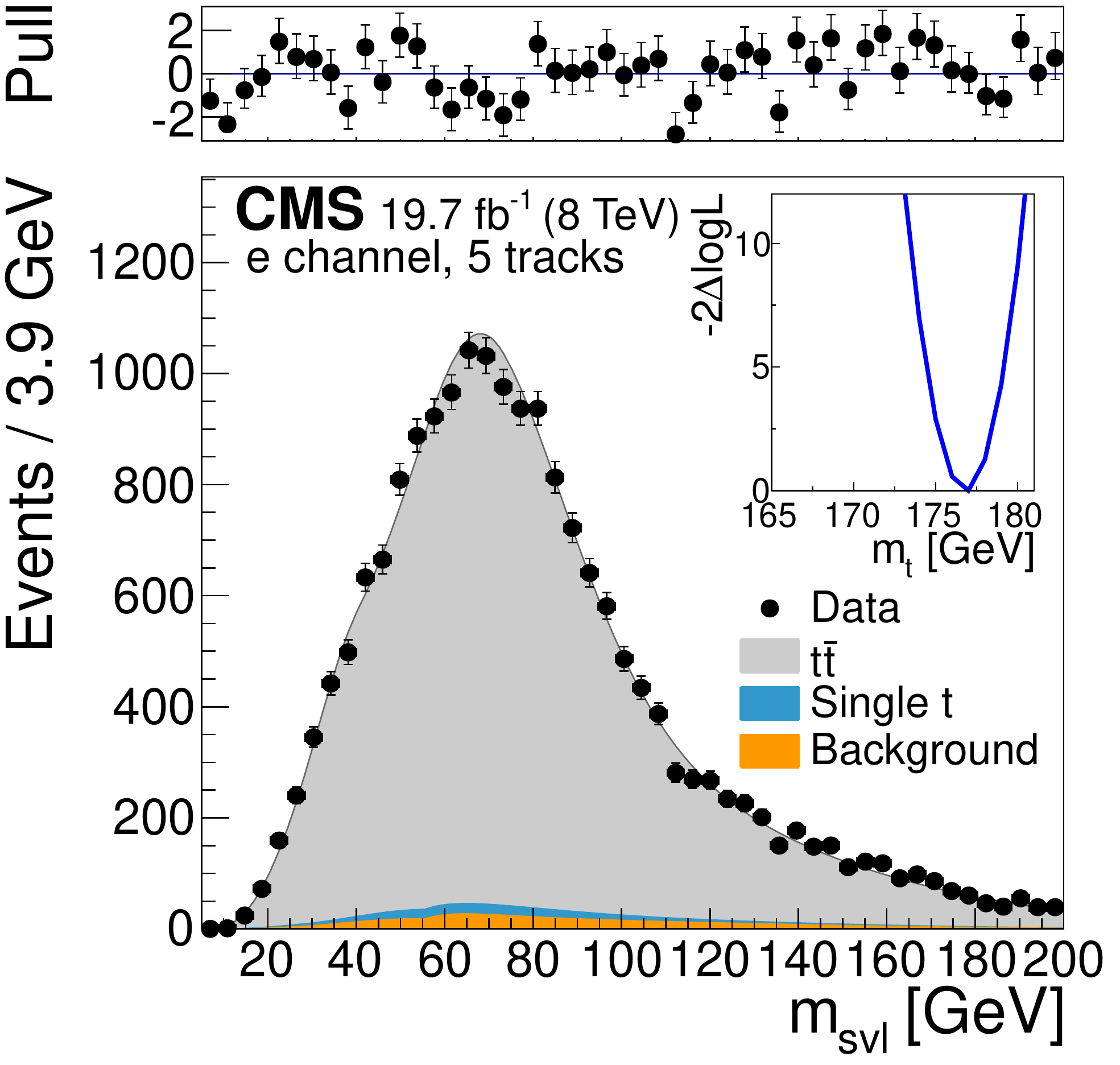}
\includegraphics[width=0.32\textwidth]{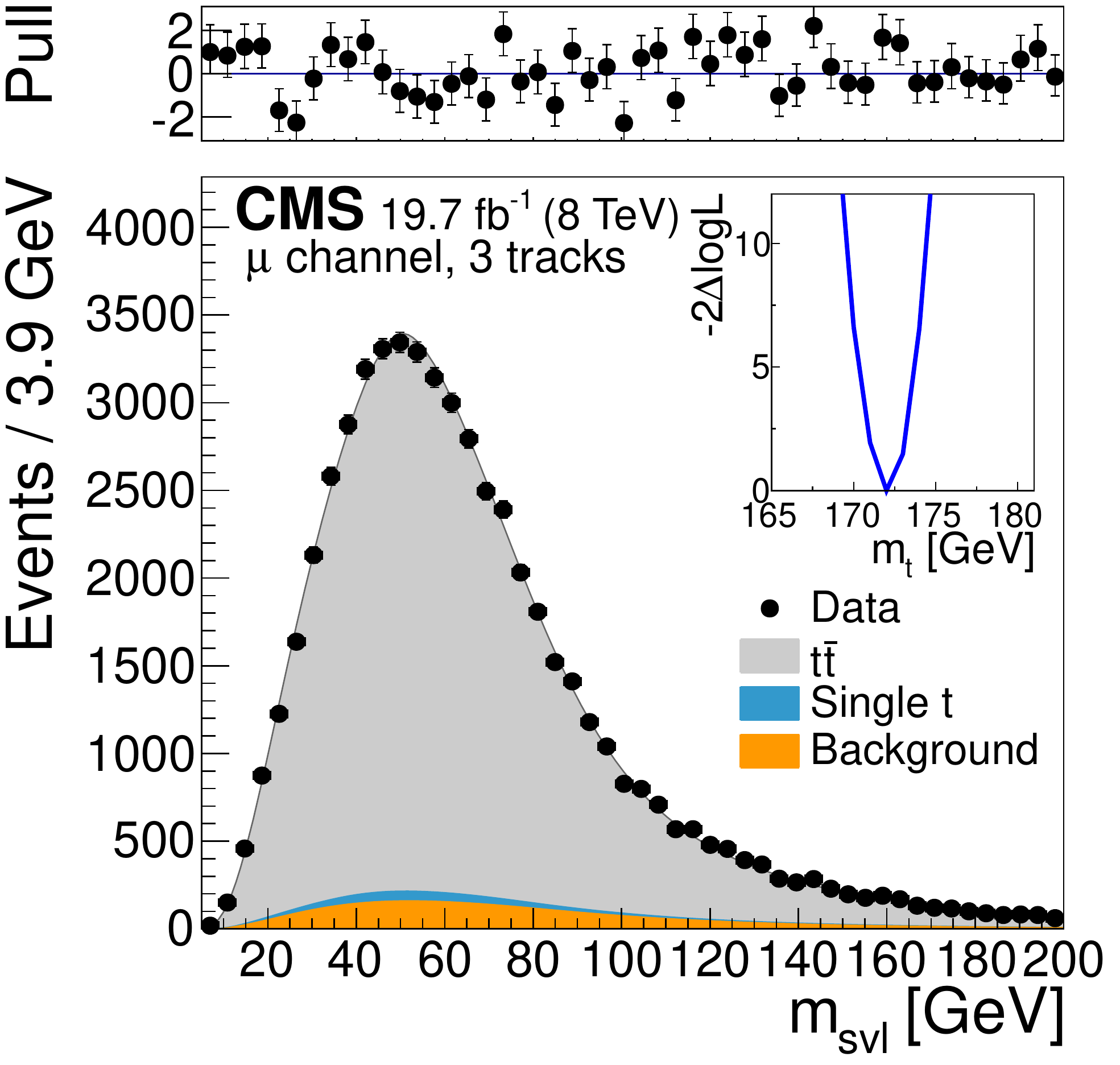}
\includegraphics[width=0.32\textwidth]{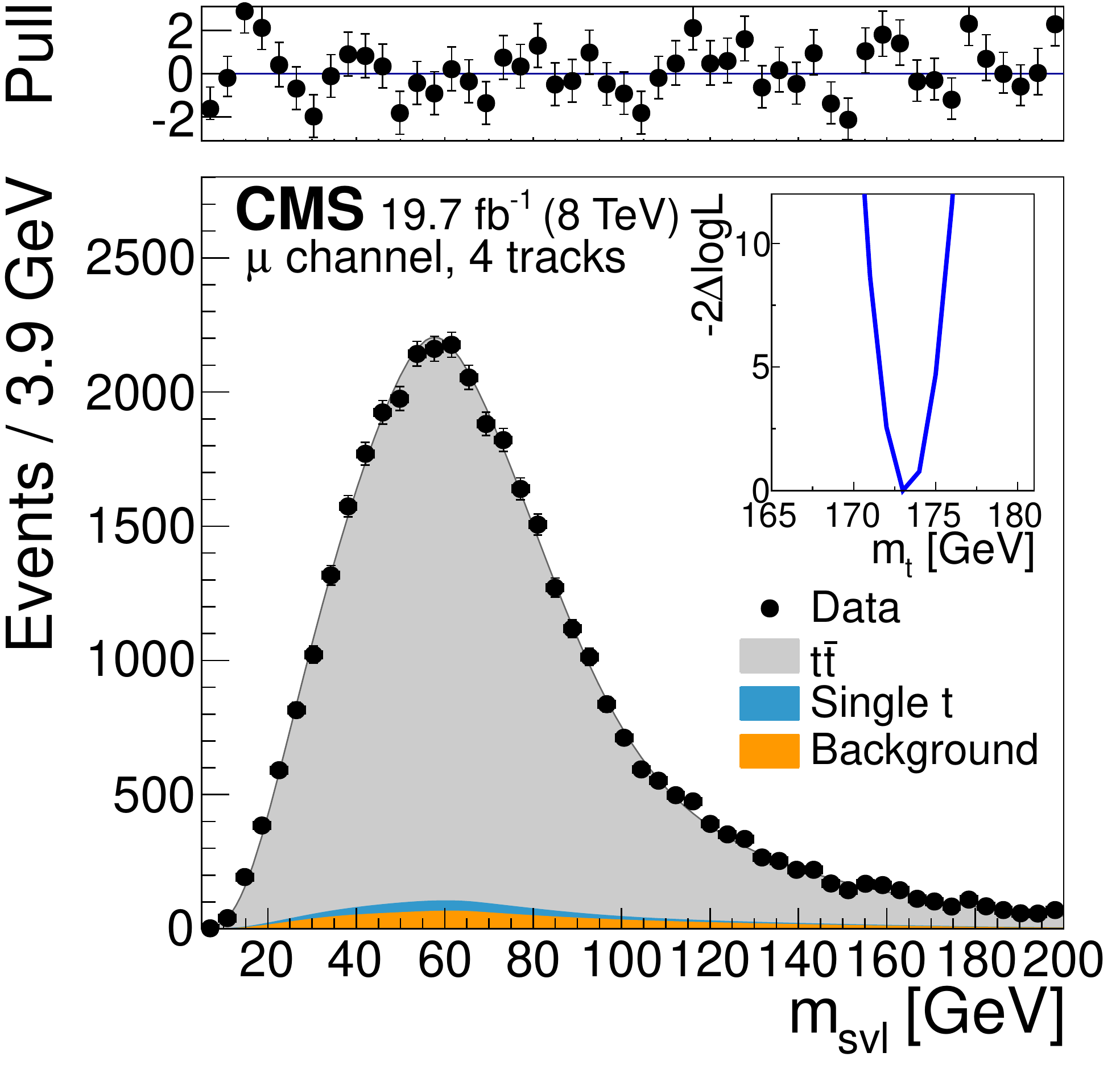}
\includegraphics[width=0.32\textwidth]{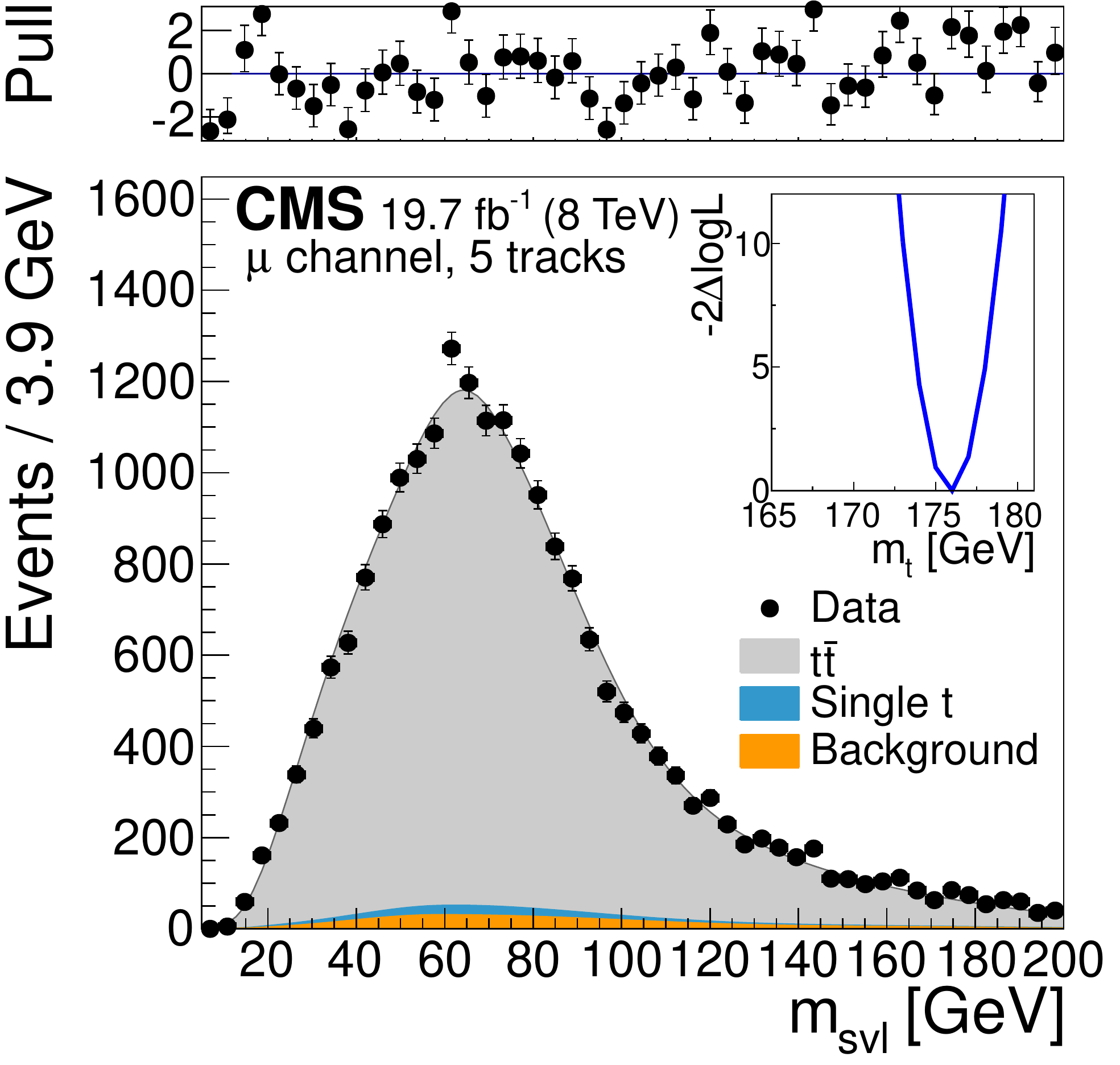}
\caption{
Template fits to the observed \msvl{} distributions for the semileptonic channels (\ejets\ and \mujets\ from top to bottom row), and for exactly three, four, and five tracks assigned to the secondary vertex (from left to right column).
The top panels show the bin-by-bin difference between the observed data and the fit result, divided by the statistical uncertainty (pull).
The inset shows the scan of the negative log-likelihood as a function of the calibrated top quark mass, accounting only for the statistical uncertainty, when performed exclusively in each event category.
}
\label{fig:msvlfitslj}
\end{figure*}

The results of the fits to the observed \msvl{} distributions in all fifteen categories are shown in Figs.~\ref{fig:msvlfitsdil} and~\ref{fig:msvlfitslj} for the dilepton and semileptonic channels, respectively.

The final results for the top quark mass are then extracted by performing a binned maximum-likelihood estimation where the observed data are compared to the expectations using Poisson statistics.
The combined likelihood is then written as:
\begin{equation*}
\ifthenelse{\boolean{cms@external}}
{
\begin{split}
\mathcal{L}&(\mtop,\mu,\vec{\theta}_{\rm bkg})\;=\;  \prod_{c=1}^{5}\,\prod_{n=3}^{5}\,\prod_{i=1}^{N_{\rm bins}} \\ & \mathcal{P}\big[N_{\rm obs}(\msvl^{i}),N_{\rm exp}(\mtop,\mu,\theta_{\rm bkg},\msvl^{i})]  \,  \mathcal{G}(0,\theta^{c,n}_{\rm bkg},0.3),
\end{split}
}
{
\mathcal{L}(\mtop,\mu,\vec{\theta}_{\rm bkg})\;=\; \prod_{c=1}^{5}\,\prod_{n=3}^{5}\,\prod_{i=1}^{N_{\rm bins}} \mathcal{P}\big[N_{\rm obs}(\msvl^{i}),N_{\rm exp}(\mtop,\mu,\theta_{\rm bkg},\msvl^{i})] \, \mathcal{G}(0,\theta^{c,n}_{\rm bkg},0.3),
}
\end{equation*}
where the products of the Poisson-distributed yields ($\mathcal{P}$) over every channel ($c$), track multiplicity category ($n$), and \msvl{} bin ($i$) are multiplied by a penalty Gaussian function for the correction of the expected background yields ($\mathcal{G}$), with a fixed width of 30\%, corresponding to the uncertainty in the background normalization.
Finally, the combined likelihood is maximized to obtain the final \mtop{} result.
The analysis has been developed using simulated events, without performing the final fit on the data until the full measurement procedure had been validated.

The method is calibrated separately in each channel and track multiplicity bin before combining them by running pseudo-experiments for each generated top quark mass point and calculating a linear calibration function from the respective extracted mass points.
Pseudo-data are generated from the combined expected shape of the top quark signals and the mixture of backgrounds with the number of generated events taken from a Poisson distribution around the expected number of events in each category.
The width of the pull distributions, \ie\ the observed bias of each fit divided by its uncertainty, indicate a proper coverage of the statistical uncertainty.
The post-calibration mass difference is below 100\MeV{} for the entire range of generated \mtop{} values, well within the statistical uncertainty of the overall measurement of 200\MeV.

\subsection{Systematic uncertainties}\label{ssec:systematics}
The size of the systematic uncertainties is evaluated from their impact on the \msvl{} shape and its propagation to the extracted \mtop{} value in the combined fit.
Modified pseudo-data are generated for each variation of the signal shape at the central mass point of 172.5\GeV, and the difference between the mass extracted from the modified data and the nominal fit is quoted as the systematic uncertainty.
The individual sources of systematic uncertainties and the determination of the shape variation are described in the following.
The final systematic uncertainties are summarized in Table~\ref{tab:systematics}.

\begin{table}[htbh!]
\begin{center}
\topcaption{Summary of the systematic uncertainties in the final measurement. In cases where there are two variations of one source of uncertainty, the first and second numbers correspond, respectively, to the down and up variations. The total uncertainties are taken as the separate quadratic sum of all positive and negative shifts. For the contributions marked with a (*), the shift of the single variation including its sign is given, but the uncertainty is counted symmetrically in both up and down directions for the total uncertainty calculation.\label{tab:systematics}}
\begin{scotch}{ll}
Source & $\Delta$\mtop{}~[$\GeV$] \\
\hline
\\[-1.5ex]
\multicolumn{2}{l}{Theoretical uncertainties}\\
\hline
$\muR/\muF$ scales \ttbar{}                          & $ +0.22 \ -\!0.20$  \\
$\muR/\muF$ scales \cPqt\ ($t$ channel)              & $ -0.04 \ -\!0.02$  \\
$\muR/\muF$ scales $ \cPqt\PW $                         & $ +0.21 \ +\!0.17$  \\
Parton shower matching scale                         & $ -0.04 \ +\!0.06$  \\
Single top quark fraction                            & $ -0.07 \ +\!0.07$  \\
Single top quark diagram interference~(*)            & $ +0.24$            \\
Parton distribution functions                        & $ +0.06 \ -\!0.04$  \\
Top quark \pt{}                                      & $ +0.82$            \\
Top quark decay width~(*)                            & $ -0.05$            \\
\cPqb\ quark fragmentation                           & $ +1.00 \ -\!0.54$  \\
Semileptonic \PB{} decays                            & $ -0.16 \ +\!0.06$  \\
\cPqb\ hadron composition~(*)                        & $ -0.09$            \\
Underlying event                                     & $ +0.07 \ +\!0.19$  \\
Color reconnection~(*)                               & $ +0.08$            \\
Matrix element generator~(*)                         & $ -0.42$            \\
$\sigma(\ttbar+{\rm heavy~flavor})$                  & $ +0.46 \ -\!0.36$  \\
\hline
Total theoretical uncertainty                        & $ +1.52 \ -\!0.86$  \\
\hline
\\[-1.5ex]
\multicolumn{2}{l}{Experimental uncertainties} \\
\hline
Jet energy scale                                     & $ +0.19 \ -\!0.17$  \\
Jet energy resolution                                & $ -0.05 \ +\!0.05$  \\
Unclustered energy                                   & $ +0.07 \ -\!0.00$  \\
Lepton energy scale                                  & $ -0.26 \ +\!0.22$  \\
Lepton selection efficiency                          & $ +0.01 \ +\!0.01$  \\
\cPqb\ tagging                                       & $ -0.02 \ -\!0.00$  \\
Pileup                                               & $ -0.05 \ +\!0.07$  \\
Secondary vertex track multiplicity~(*)                   & $ -0.06$            \\
Secondary vertex mass modeling~(*)                        & $ -0.29$            \\
Background normalization                             & $ {<}0.03$  \\
\hline
Total experimental uncertainty                       & $ +0.43 \ -\!0.44$  \\
\hline
\\[-1.5ex]
Total systematic uncertainty                  & $+1.58 \ -\!0.97$ \\
Statistical uncertainty                        & ${\pm}0.20$  \\
\end{scotch}
\end{center}
\end{table}

\subsubsection{Modeling and theoretical uncertainties}\label{sssec:theosysts}

\begin{itemize}

\item {\bf Choice of renormalization and factorization scales:}
\sloppy{
The factorization and renormalization scales used in the signal simulation are set to a common value, $Q$, defined by \mbox{$Q^2=\mtop^2 + \sum{(\pt^\text{parton})}^2$}, where the sum runs over all extra partons in the event.
Two alternative data sets with a variation $\muR=\muF=2Q$ or $Q/2$ are used to estimate the systematic effect from the choice of scales.
These variations are observed to provide a conservative envelope of the additional jet multiplicity observed in data~\cite{Khachatryan:2015mva}.
The scale choice for single top quark $t$ and $ \cPqt\PW $~channels has a smaller effect on the measurement because the production happens through an electroweak interaction and because single top quark events only make up about 5\% of the total yield.
Dedicated single top quark data samples with \muF\ and \muR\ varied by a factor $2$ or $1/2$ are generated and used to estimate the effect.
}

\item {\bf Matrix element to parton shower matching scale:}
The choice of the threshold in the event generation at which additional radiation is produced by the \PYTHIA{} showering instead of matrix element calculations in \MADGRAPH{} is expected to have a small impact on the shape of \msvl, affecting mostly the ``unmatched'' lepton-SV pairings, which constitute only about 5\% of the total.
Variations of this threshold are furthermore observed to have small impact on the kinematic properties of extra jets~\cite{Khachatryan:2015mva}.
The effect is estimated using dedicated samples with the nominal threshold (20\GeV) varied up and down by a factor of 2.

\item {\bf Single top quark fraction:}
The overall signal shapes in each category are constructed from \ttbar{} events and events from single top quark production, with their relative fractions fixed to the expectation from theory.
Because of a relative difference in their respective shapes, a deviation in this fraction can have an impact on the final mass measurement.
The effect is estimated by repeating the fits with the relative fraction of single top quark events in the signal shape varied by $\pm20\%$.
The size of the variation reflects the experimental uncertainty in the overall cross section of single top quark production~\cite{Chatrchyan:2014tua,Khachatryan:2014iya}.

\item {\bf Single top quark interference:}
Interference between \ttbar\ pair production and single top quark production in the $ \cPqt\PW $ channel at next-to-leading order in QCD is resolved in the $ \cPqt\PW $ signal generation by removing all doubly-resonant diagrams in the calculation~\cite{Frixione:2008yi,Belyaev:1998dn,White:2009yt}.
A different scheme for the resolution of the diagram interference can be defined where a gauge-invariant subtraction term modifies the $ \cPqt\PW $ cross section to cancel the contributions from \ttbar.
Samples using the second scheme are generated and compared and the difference is quoted as a systematic uncertainty~\cite{Tait:1999cf,Frixione:2008yi}.

\item {\bf Parton distribution functions:}
Uncertainties from the modeling of parton momentum distributions inside the incoming protons are evaluated using the diagonalized uncertainty sources of the CT10 PDF set~\cite{Pumplin:2002vw}.
Each source is used to derive event-by-event weights, which are then applied to obtain a variation of the signal \msvl{} shape.
The maximal difference with respect to the nominal signal sample is quoted as the systematic uncertainty.

\item {\bf Top quark \pt{} modeling:}
Measurements of the differential \ttbar{} production cross section reveal an observed top quark \pt{} spectrum that is softer than what is predicted from simulation~\cite{Khachatryan:2015oqa}.
The difference between the unfolded data and the simulation based on \MADGRAPH{} is parametrized and can be used to calculate event-by-event weights correcting the spectrum.
This reweighting is not applied when calibrating the measurement, as it introduces a dependence on the true top quark mass.
The impact of the full difference between the predicted spectrum used in the calibration (at \mtop=172.5\GeV) and the data-corrected spectrum is estimated by comparing the result from reweighted pseudo-data to the nominal value.
The difference is then added as a one-sided systematic uncertainty in the extracted mass value.
The effect of the reweighting on the simulated \msvl{} shape for correct and wrong lepton-vertex pairings is shown in Fig.~\ref{fig:systvariations}.

\item {\bf Top quark decay width:}
The decay width of the top quark has been experimentally determined with a precision of about 10\%~\cite{Khachatryan:2014nda}.
A dedicated sample with an increased width is used to estimate the impact on the mass measurement, and the difference is quoted as an uncertainty.

\item {\bf \cPqb{} quark fragmentation:}
A variation in the momentum transfer from \cPqb{} quark to $\PQb$~hadron has a direct impact in the \msvl\ distribution, and correspondingly, the uncertainty from the used \cPqb\ quark fragmentation function on the extracted top quark mass is expected to be significant.
As shown in Section~\ref{sec:modeling}, the average momentum transfer in the nominal \PYTHIA{} \ztwostar{} tune is found to be significantly softer than that seen in \ttbar\ events in the data, whereas the \ztsrblep\ variation that follows a fragmentation function measured at LEP is in better agreement.
Its soft and hard variations provide one standard deviation variations of the shape parameters, and are used to estimate the systematic uncertainty.
Variations of the \msvl{} shape for the central \ztsrblep\ fragmentation function, its soft and hard variations, as well as the nominal \ztwostar\ fragmentation are shown in Fig.~\ref{fig:systvariations}.
The impact of the choice of \cPqb\ quark fragmentation function on the extracted top quark mass is shown in Fig.~\ref{fig:mtopvsfrag}.
To first order the measured \mtop{} value depends only on the average momentum transfer, as indicated by the linear dependence on $\langle\pt(\PB)/\pt(\cPqb)\rangle$.
The extracted mass changes by about 0.61\GeV{} for each percent of change in the average momentum transfer.

\begin{figure}[htp]
	\centering
	\includegraphics[width=0.45\textwidth]{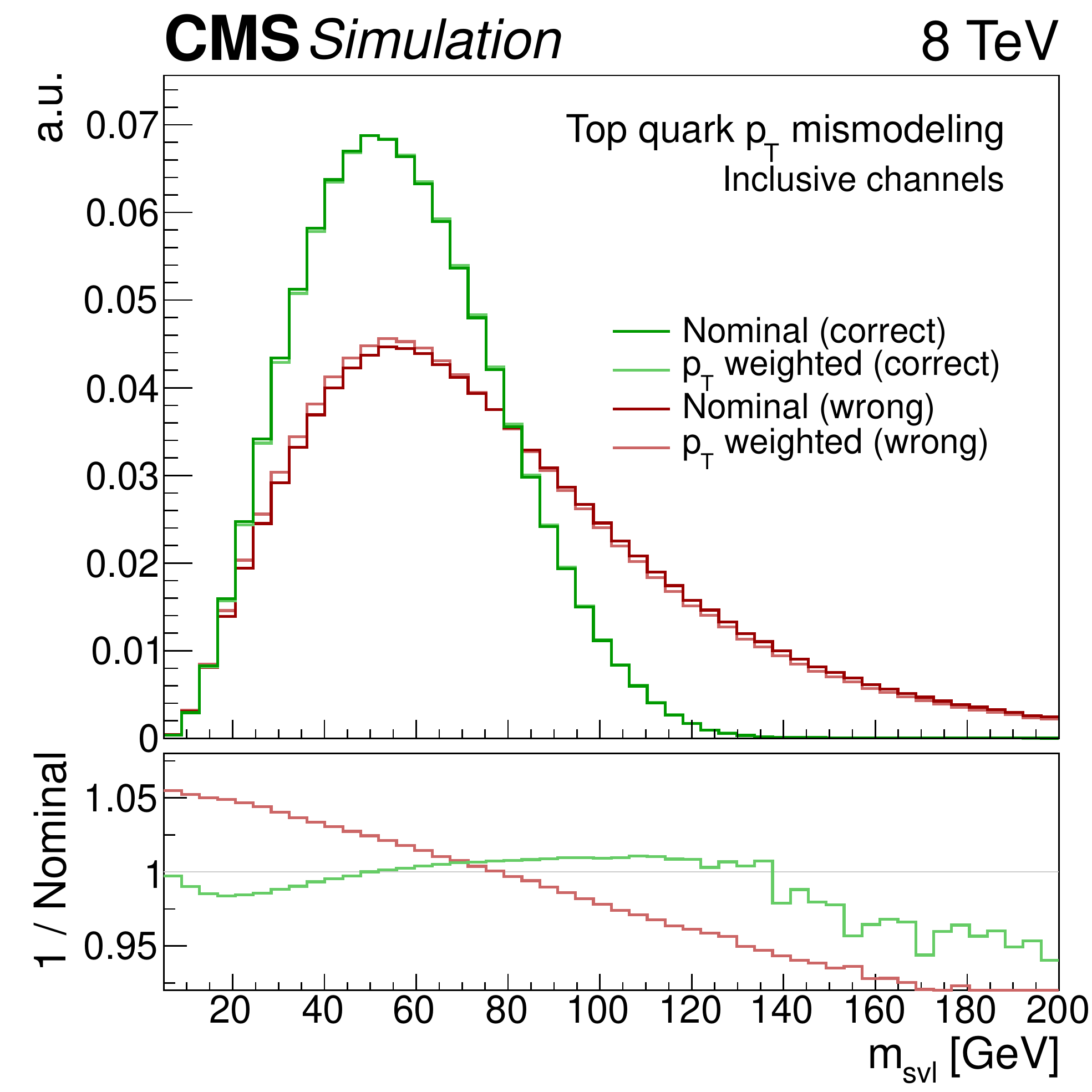}
	\includegraphics[width=0.45\textwidth]{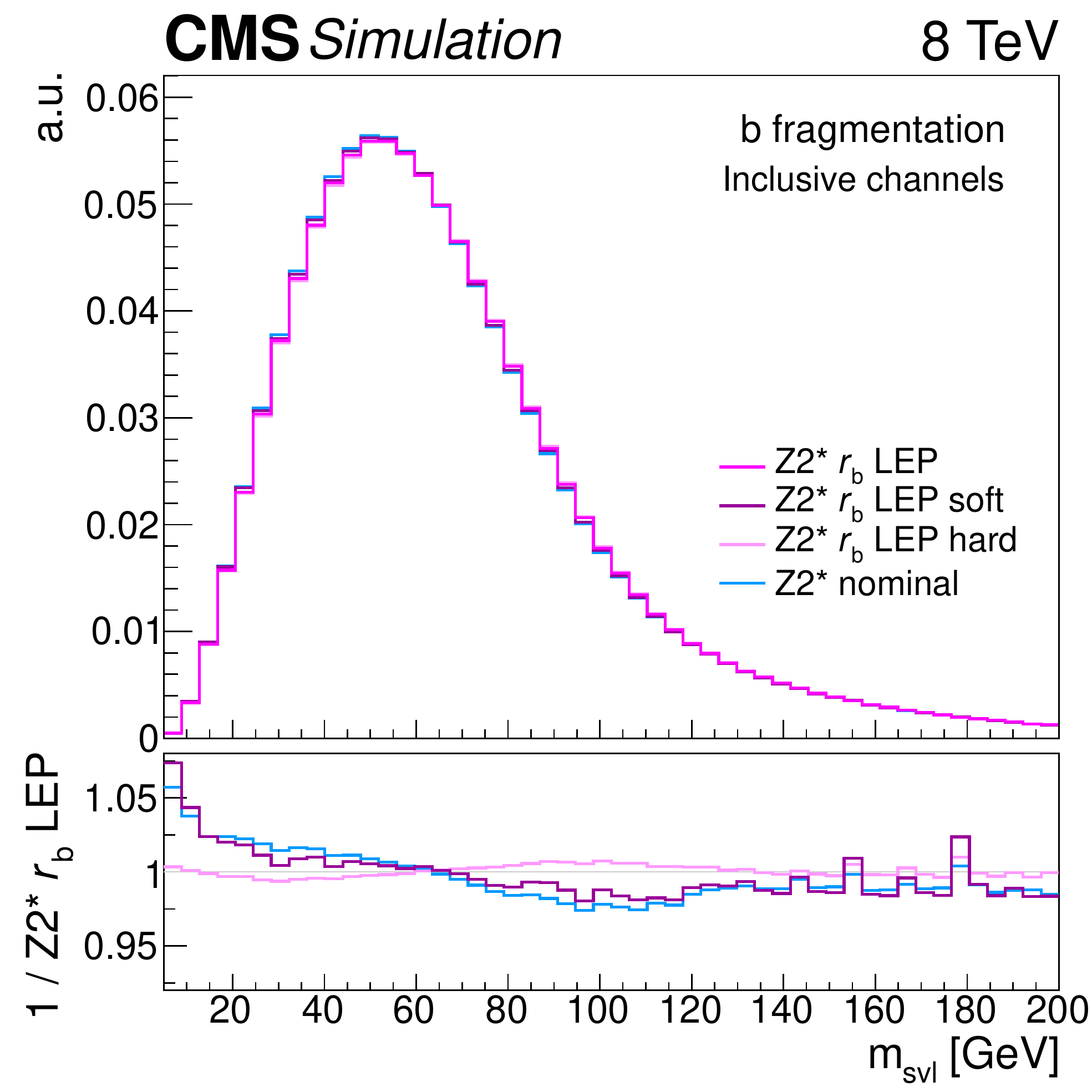}
	\caption{
	Variation of the simulated \msvl{} shape with systematic effects: reweighting of the simulated top quark \pt{} shape to the observed distribution, separately for correct and wrong lepton-vertex pairings (\cmsLeft); and different \cPqb\ quark fragmentation function shapes (\cmsRight).
	The bottom panels in the two plots show the ratios between the top quark \pt{} reweighted and nominal shapes for the correct and wrong pairings (\cmsLeft), and between the various fragmentation models and the central \ztsrblep\ tune (\cmsRight).
	}
	\label{fig:systvariations}
\end{figure}

\begin{figure}[htp]
  \centering
  \includegraphics[width=0.45\textwidth]{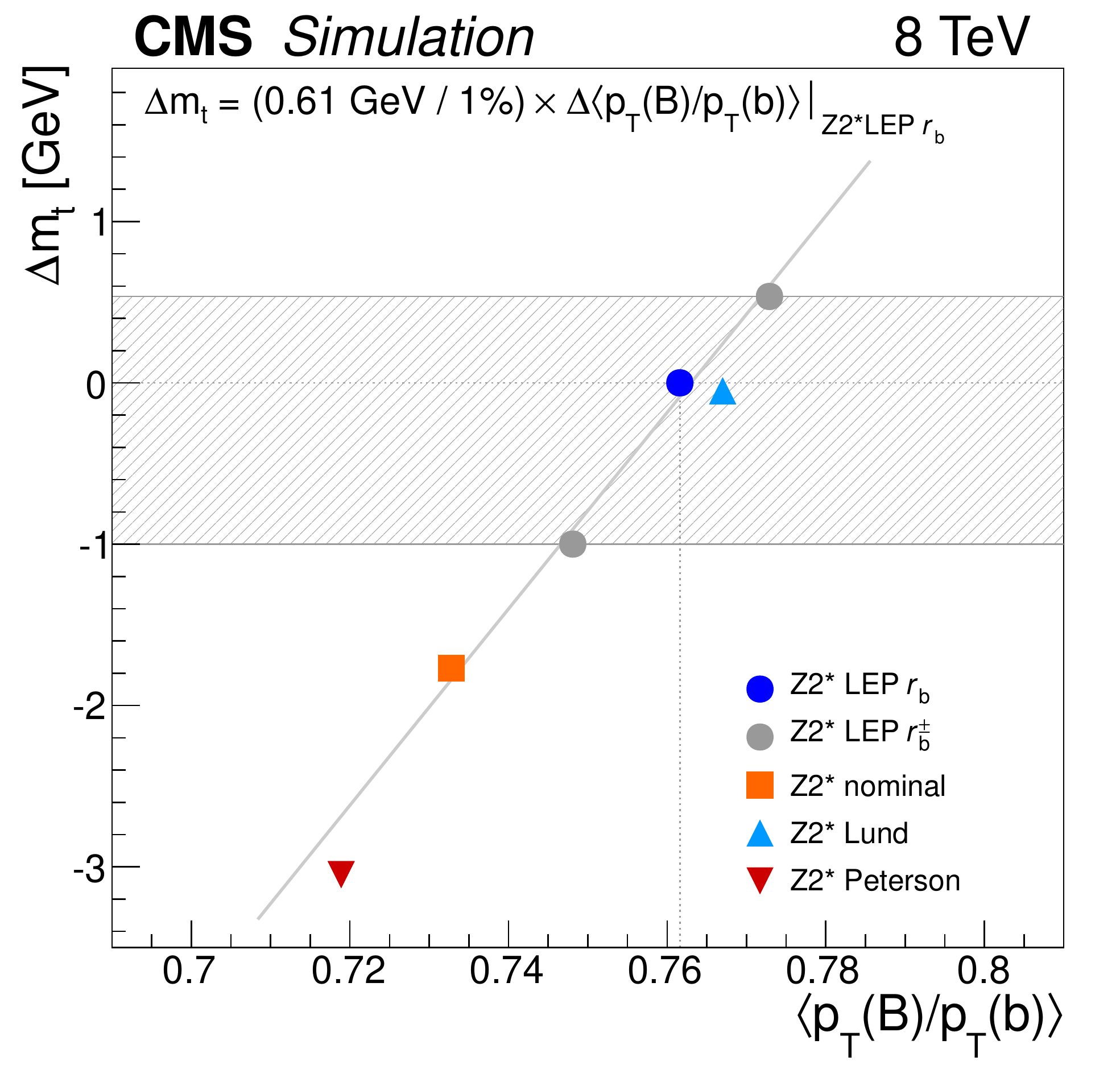}
  \caption{
    Impact of the average \cPqb\ quark fragmentation, $\langle\pt(\PB)/\pt(\cPqb)\rangle$, on the extracted \mtop\ value, for various different fragmentation shapes.
    The horizontal band represents the contribution of the \cPqb{} quark fragmentation model to the systematic uncertainty in the measurement of the top quark mass, which is estimated from the \ztsrblep~$\!^{\pm}$ variations.
    A linear fit to the effects on the different variations (the line in the figure) suggests a relative change in the measured top quark mass of 0.61\GeV\ for each percent change in average momentum transfer.
  }\label{fig:mtopvsfrag}
\end{figure}

\item {\bf Semileptonic \PB~meson branching fractions:}
The effect of the uncertainties in semileptonic $\PQb$~hadron branching fractions is estimated by varying the fraction of \cPqb{} jets containing neutrinos down by 0.45\% and up by 0.77\%, covering the uncertainties in the experimentally measured semileptonic branching fractions of \PBz{} and \PBpm{} mesons~\cite{Agashe:2014kda}.

\item {\bf \cPqb\ hadron composition:}
The \PYTHIA{} \ztwostar{} tune produces an average composition of about 40\%~\PBz{}, 40\%~\PBpm{}, 12\%~\PBs{}, and 8\% heavier $\PQb$~hadron states in the hadronization of \cPqb{} quarks.
An improved version of this tune that takes into account hadron multiplicity measurements~\cite{Agashe:2014kda} is used to estimate the uncertainty due to the composition of $\PQb$~hadrons in the \cPqb{} jets.

\item {\bf Hadronization model cross-check:}
To test for additional uncertainties arising from the usage of the Lund string hadronization model in \PYTHIA~\cite{Andersson:1983ia} in the default simulation, additional cross-checks are performed with alternative hadronization models as used in \HERWIG{}.
However, an inclusive comparison of the two parton shower and hadronization frameworks entangles various different effects in an incoherent and nontransparent manner and includes uncertainties that are already evaluated in dedicated studies in more sound ways.
The inclusive \PYTHIA{}-\HERWIG{} difference is therefore not included as a systematic uncertainty.
Evaluating whether there are indeed additional sources of uncertainty arising when comparing different hadronization models requires a comparison without changing the parton shower model, the hard-scattering simulation, or the \cPqb\ quark fragmentation functions.
Such a check is possible in the \SHERPA{} 2.1.0 framework~\cite{Gleisberg:2008ta}, which permits a \pt-ordered parton shower model to be used, interfaced with a cluster hadronization model as used in \HERWIG{} or with the Lund string model of \PYTHIA{}.
The change in hadronization model entails a difference in hadron flavor multiplicities, with the cluster model tending to yield more heavy $\PB_{\rm c}$ mesons and $\Lambda_{\cPqb}$ baryons.
Restricting the study to the dominant production of \PBz{} and \PBpm{} mesons reveals a different \cPqb\ quark fragmentation function shape between the two models.
As the uncertainty from this effect is already covered by a more extreme variation in the dedicated \cPqb\ quark fragmentation uncertainty, the distributions are reweighted to a common \cPqb\ parton to \cPqb\ hadron momentum transfer distribution to remove any difference in fragmentation shapes.
The resulting lepton + \cPqb{} jet invariant mass distributions for cluster and Lund string fragmentation are found to be in very good agreement and do not warrant any additional uncertainty in the top quark mass measurement.

\item {\bf Underlying event and color reconnection:}
Effects from the modeling of the proton collision remnants and multiparton interactions (the underlying event) and from the color connection of the \cPqb{} quark fragmentation products to the rest of the event (color reconnection) are estimated using dedicated samples with variations of the Perugia 11 (P11) underlying event tunes~\cite{Skands:2010ak}.
Two variations, one with altered multiparton interactions and one based on the Tevatron data are used to evaluate the effect of the underlying event modeling.
A separate sample, in which color reconnection effects are not simulated, is used to gauge the impact from the modeling of this effect.
In both cases, the full difference of the results obtained on the modified samples and the case of using pseudo-data from the central P11 tune are quoted as the systematic uncertainty.

\item {\bf Matrix element generator:}
The default Born-level matrix element generator, \MADGRAPH{}, is substituted by a \POWHEG{} simulation based on the heavy-quark pair production (hvq) model~\cite{Frixione:2007nu} at NLO accuracy for \ttbar{} production and at leading order for the top quark decays.
In both cases, the matrix element generators are interfaced with \PYTHIA{} for parton showering.
The difference, propagated to the mass measurement, is reported as a systematic uncertainty.

Furthermore, the effect of including NLO corrections in the modeling of the top quark decay is studied using the parton-level \MCFM\ program~\cite{Campbell:2010ff, Campbell:2012uf}.
Since no fragmentation or parton shower evolution is included in the simulation and therefore the actual impact on the mass measurement is uncertain, the result is only reported here but not included as a systematic uncertainty.
By reweighting the mass of the lepton-\cPqb-jet system generated by \MADGRAPH\ to the differential cross sections predicted by \MCFM, with and without applying NLO corrections to the top quark decay, a +1.29\GeV\ shift in the calibrated mass in the $\emu$ channel is observed.

\item {\bf Modeling of the associated production of \ttbar\ with heavy flavors:}
While the simulation is observed to describe the shape of the different distributions for \ttbar\kern -3pt+heavy flavors well (most notably \ttbar\kern -3pt+\bbbar), these predictions tend to underestimate the total cross section~\cite{Aad:2015yja,Khachatryan:2015mva}.
To evaluate the impact on the measurement, the nominal simulation is compared to the one obtained after reweighting the contribution from extra \cPqb\ jets in the simulation by the data-to-theory scale factor measured in~\cite{Khachatryan:2015mva}.
A symmetric variation of the expected extra heavy-flavor content is used to estimate this uncertainty.

\end{itemize}

\subsubsection{Experimental uncertainties}\label{sssec:expsysts}
\begin{itemize}

\item {\bf Jet energy scale and jet energy resolution:}
By design, the reconstructed jet energy does not affect the \msvl{} observable.
However jet momenta are used in the event selection and therefore variations of the jet energy have an effect on the event yields that enter the bins of the \msvl{} distributions.
The effects are estimated by rescaling the reconstructed jet energies depending on \pt{} and $\eta$ before performing the event selection.
The effect of jet energy resolution on the measured distributions is estimated by inflating or deflating the resolution within the measured uncertainties and propagating the effects to the final distributions.
The varied \msvl{} distributions are used to generate pseudo-data, and the full differences to the nominal sample are quoted as the systematic uncertainties.

\item {\bf Unclustered energy:}
The missing transverse energy is used only in the event selection for the \ee{} and \mumu{} channels to suppress events containing neutrinoless \cPZ~boson decays.
Since the DY yield is normalized from a dedicated data control region, the effect from the \MET{} resolution is expected to be small.
It is estimated by varying the amount of energy that is not clustered into jets in the \MET{} calculation by ${\pm}10\%$ and studying its impact on the observed \msvl{} distributions.

\item {\bf Lepton momentum scale:}
The reconstructed lepton momenta directly affect the \msvl{} spectrum.
The uncertainty in the measured energy scale for electrons depends on \pt{} and $\eta$ and varies between about 0.6\% in the central barrel region and about 1.5\% in the forward region~\cite{Khachatryan:2015hwa}.
The muon momentum scale is known within an uncertainty of about 0.2\%~\cite{Chatrchyan:2013sba}.
Varying the scales up and down within their measured uncertainties---as a function of \pt{} and $\eta$ for electrons---produces a shift in the \msvl{} distribution that is propagated to the final mass measurement and quoted as a systematic uncertainty.

\item {\bf Lepton selection efficiency:}
Similar to the jet energy scales, the requirements applied when selecting lepton candidates for the analysis affect the event yields in the \msvl{} distributions and can cause a slight change in the extracted top quark mass.
The measured electron and muon selection efficiencies are varied within their uncertainties and the difference is quoted as a systematic uncertainty.

\item {\bf \cPqb{} tagging efficiency and mistag rate:}
The \ttbar{} event selection relies on the use of a \cPqb{} tagging algorithm to select jets originating from the hadronization of a \cPqb{} quark.
The impact on \msvl{} from the uncertainties in the signal and background efficiencies of this algorithm are estimated by varying the efficiencies within their measured uncertainties and propagating the effect to the final result.

\item {\bf Pileup:}
The effect of additional concurrent \Pp\Pp{} interactions on the measured precision is estimated by varying the cross section for inelastic \Pp\Pp{} collisions used in the pileup generation by ${\pm}5\%$, and propagating the difference to the extracted \mtop{} result.

\item {\bf Secondary-vertex track multiplicity:}
The distribution of the number of tracks assigned to secondary vertices is not well described by simulation, as has been observed in several processes involving \cPqb{} quarks.
Generally, the data shows about 5--10\% fewer tracks than the simulation.
As the analysis is carried out in exclusive bins of track multiplicity to minimize the impact of this issue, it only enters as a second-order effect when combining the results from different bins, as the individual bins would be assigned slightly different weights in simulation.
This is corrected for by reweighting each bin content by the yield observed in the data, and the impact of this reweighting on the final result is quoted as a remaining systematic uncertainty.

\item {\bf Secondary-vertex mass modeling:}
A discrepancy between the observed secondary vertex mass (\ie\ the invariant mass of the tracks used to reconstruct the vertex) and the one predicted in the simulation is observed.
The effect is propagated in the \msvl{} shape by weighting the simulated events to reflect the observed distributions in each bin of track multiplicity, and the resulting shift in the extracted top quark mass is quoted as a systematic uncertainty.

\item {\bf Background normalization:}
Processes not involving top quarks constitute about 5\% of the overall selected events and their combined yield is allowed to float within about 30\% in the fit.
The normalization of the main background processes is furthermore determined in dedicated control samples in the data.
To estimate the uncertainty in the result stemming from the uncertainty in the background normalization, the expected yields of backgrounds are varied within their uncertainties, and the resulting change in the \msvl{} shape is propagated to the final result.
These variations are observed to have a negligible impact on the measurement
as they are absorbed by upward/downward variations of the background yields in the fit.

\end{itemize}

\subsection{Results}\label{ssec:results}
\begin{figure*}[htp]
\centering
\includegraphics[width=0.90\textwidth]{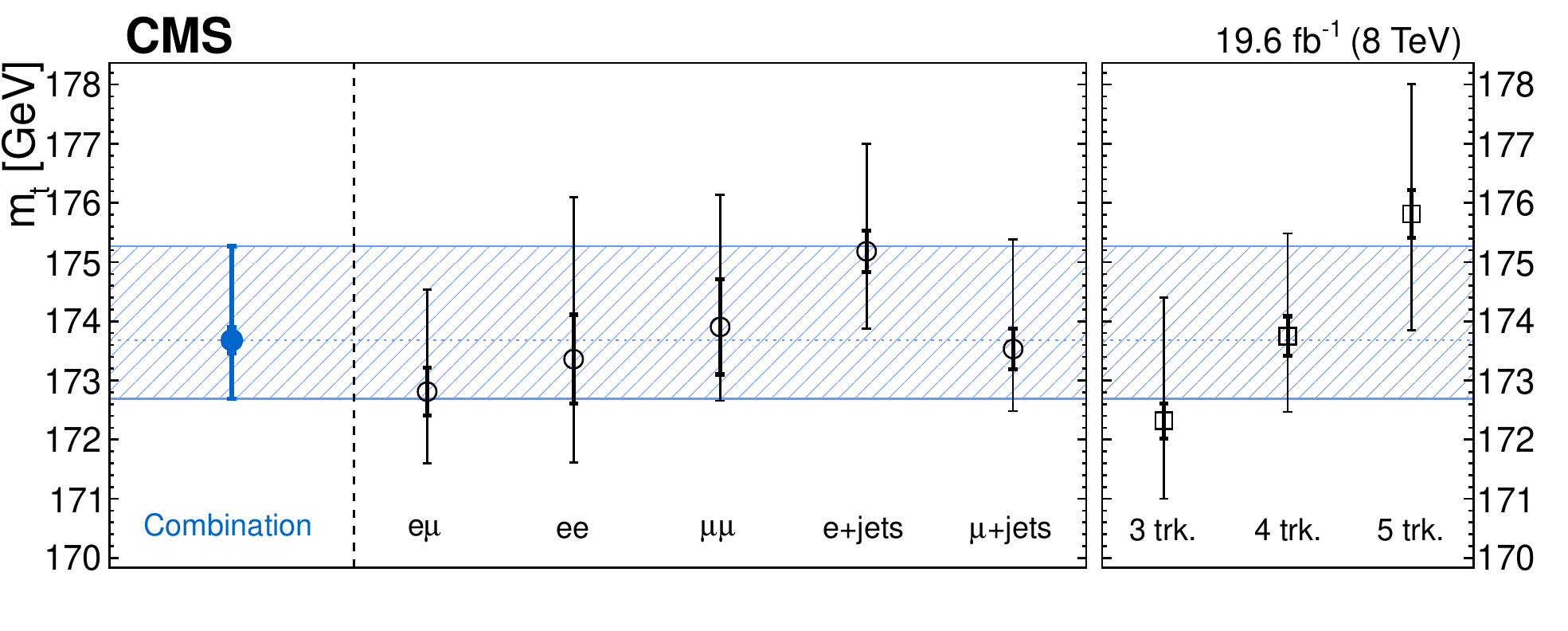}
\caption{
Results of the \mtop{} measurement in the individual channels and their combination.
Smaller and thicker error bars show the statistical uncertainty, whereas the thinner bars show the combined statistical and systematic uncertainty.
The right panel shows the extracted mass when performing the analysis in separate track multiplicity bins, combining the lepton channels.
}
\label{fig:results}
\end{figure*}

The top quark mass is measured from the invariant mass distribution of leptons and reconstructed secondary vertices from $\PQb$~hadron decays using only charged particles.
After calibrating the measurement with simulated events, a value of
\begin{equation*}
\mtop=173.68 \ \pm 0.20 {\rm (stat)} \ ^{+1.58}_{-0.97}{\rm (syst)}\, \GeV
\end{equation*}
is obtained from the data, with a combined uncertainty of $^{+1.59}_{-\!0.99}\GeV$.
The overall systematic uncertainty is dominated by the uncertainty in the \cPqb\ quark fragmentation and the modeling of kinematic properties of top quarks with minimal sensitivity to experimental uncertainties.
Figure~\ref{fig:results} shows the combined result as well as the values obtained separately for the five lepton channels and the three track multiplicity bins.
The observed trend as a function of the track multiplicity is compatible with the results obtained regarding the modeling of the relative momentum of secondary vertices inside jets, as discussed in Section~\ref{sec:modeling}.

\section{Summary and prospects}\label{sec:conclusions}
A novel measurement of the top quark mass has been presented, using an observable that relies entirely on the reconstruction of charged particles.
It shows minimal sensitivity to experimental sources of uncertainty.
The final result yields a value of $\mtop=173.68^{+1.59}_{-0.99}\GeV$, equivalent to a precision of well below one percent.
The overall uncertainty is dominated by the \cPqb\ quark fragmentation modeling uncertainty of $+1.00/\!\!-\!0.54\GeV$ and the uncertainty in the modeling of the top quark \pt{} of $+0.82\GeV$.
Experimental uncertainties related to the understanding of jet energy scales only affect the event acceptance and are therefore virtually irrelevant to the final result.
Studies of the \cPqb\ quark fragmentation with reconstructed secondary vertices and inclusively reconstructed charm quark mesons are used to select the central \cPqb\ quark fragmentation shape and to validate the systematic uncertainty.

With the significantly larger data sets becoming available for analysis from the current 13\TeV\ run of the LHC, this method could be extended to constrain the \cPqb\ quark fragmentation, using the properties of the secondary vertices or charmed mesons, while measuring the top quark mass.
This is expected to lead to a significant reduction of the overall uncertainty.
Furthermore, theoretical uncertainties related to kinematic properties of top quarks and scale choices in QCD calculations are expected to decrease with the next generation of Monte Carlo event generators.

Finally, this result is complementary to standard measurements relying on kinematic properties of jets.
The precision of such analyses is typically limited by the uncertainty from the modeling of hadronization effects, influencing the understanding of the jet energy scale, while not much affected by the choice of \cPqb\ quark fragmentation model and the modeling of top quark kinematic properties.
Therefore, a combination of this result with standard measurements could optimally benefit from independent sources of systematic uncertainties.

\section*{Acknowledgments}
\hyphenation{Bundes-ministerium Forschungs-gemeinschaft Forschungs-zentren} We congratulate our colleagues in the CERN accelerator departments for the excellent performance of the LHC and thank the technical and administrative staffs at CERN and at other CMS institutes for their contributions to the success of the CMS effort. In addition, we gratefully acknowledge the computing centers and personnel of the Worldwide LHC Computing Grid for delivering so effectively the computing infrastructure essential to our analyses. Finally, we acknowledge the enduring support for the construction and operation of the LHC and the CMS detector provided by the following funding agencies: the Austrian Federal Ministry of Science, Research and Economy and the Austrian Science Fund; the Belgian Fonds de la Recherche Scientifique, and Fonds voor Wetenschappelijk Onderzoek; the Brazilian Funding Agencies (CNPq, CAPES, FAPERJ, and FAPESP); the Bulgarian Ministry of Education and Science; CERN; the Chinese Academy of Sciences, Ministry of Science and Technology, and National Natural Science Foundation of China; the Colombian Funding Agency (COLCIENCIAS); the Croatian Ministry of Science, Education and Sport, and the Croatian Science Foundation; the Research Promotion Foundation, Cyprus; the Ministry of Education and Research, Estonian Research Council via IUT23-4 and IUT23-6 and European Regional Development Fund, Estonia; the Academy of Finland, Finnish Ministry of Education and Culture, and Helsinki Institute of Physics; the Institut National de Physique Nucl\'eaire et de Physique des Particules~/~CNRS, and Commissariat \`a l'\'Energie Atomique et aux \'Energies Alternatives~/~CEA, France; the Bundesministerium f\"ur Bildung und Forschung, Deutsche Forschungsgemeinschaft, and Helmholtz-Gemeinschaft Deutscher Forschungszentren, Germany; the General Secretariat for Research and Technology, Greece; the National Scientific Research Foundation, and National Innovation Office, Hungary; the Department of Atomic Energy and the Department of Science and Technology, India; the Institute for Studies in Theoretical Physics and Mathematics, Iran; the Science Foundation, Ireland; the Istituto Nazionale di Fisica Nucleare, Italy; the Ministry of Science, ICT and Future Planning, and National Research Foundation (NRF), Republic of Korea; the Lithuanian Academy of Sciences; the Ministry of Education, and University of Malaya (Malaysia); the Mexican Funding Agencies (CINVESTAV, CONACYT, SEP, and UASLP-FAI); the Ministry of Business, Innovation and Employment, New Zealand; the Pakistan Atomic Energy Commission; the Ministry of Science and Higher Education and the National Science Center, Poland; the Funda\c{c}\~ao para a Ci\^encia e a Tecnologia, Portugal; JINR, Dubna; the Ministry of Education and Science of the Russian Federation, the Federal Agency of Atomic Energy of the Russian Federation, Russian Academy of Sciences, and the Russian Foundation for Basic Research; the Ministry of Education, Science and Technological Development of Serbia; the Secretar\'{\i}a de Estado de Investigaci\'on, Desarrollo e Innovaci\'on and Programa Consolider-Ingenio 2010, Spain; the Swiss Funding Agencies (ETH Board, ETH Zurich, PSI, SNF, UniZH, Canton Zurich, and SER); the Ministry of Science and Technology, Taipei; the Thailand Center of Excellence in Physics, the Institute for the Promotion of Teaching Science and Technology of Thailand, Special Task Force for Activating Research and the National Science and Technology Development Agency of Thailand; the Scientific and Technical Research Council of Turkey, and Turkish Atomic Energy Authority; the National Academy of Sciences of Ukraine, and State Fund for Fundamental Researches, Ukraine; the Science and Technology Facilities Council, UK; the US Department of Energy, and the US National Science Foundation.

Individuals have received support from the Marie-Curie program and the European Research Council and EPLANET (European Union); the Leventis Foundation; the A. P. Sloan Foundation; the Alexander von Humboldt Foundation; the Belgian Federal Science Policy Office; the Fonds pour la Formation \`a la Recherche dans l'Industrie et dans l'Agriculture (FRIA-Belgium); the Agentschap voor Innovatie door Wetenschap en Technologie (IWT-Belgium); the Ministry of Education, Youth and Sports (MEYS) of the Czech Republic; the Council of Science and Industrial Research, India; the HOMING PLUS program of the Foundation for Polish Science, cofinanced from European Union, Regional Development Fund; the OPUS program of the National Science Center (Poland); the Compagnia di San Paolo (Torino); MIUR project 20108T4XTM (Italy); the Thalis and Aristeia programs cofinanced by EU-ESF and the Greek NSRF; the National Priorities Research Program by Qatar National Research Fund; the Rachadapisek Sompot Fund for Postdoctoral Fellowship, Chulalongkorn University (Thailand); and the Welch Foundation, contract C-1845.

\clearpage

\bibliography{auto_generated}

\cleardoublepage \appendix\section{The CMS Collaboration \label{app:collab}}\begin{sloppypar}\hyphenpenalty=5000\widowpenalty=500\clubpenalty=5000\textbf{Yerevan Physics Institute,  Yerevan,  Armenia}\\*[0pt]
V.~Khachatryan, A.M.~Sirunyan, A.~Tumasyan
\vskip\cmsinstskip
\textbf{Institut f\"{u}r Hochenergiephysik der OeAW,  Wien,  Austria}\\*[0pt]
W.~Adam, E.~Asilar, T.~Bergauer, J.~Brandstetter, E.~Brondolin, M.~Dragicevic, J.~Er\"{o}, M.~Flechl, M.~Friedl, R.~Fr\"{u}hwirth\cmsAuthorMark{1}, V.M.~Ghete, C.~Hartl, N.~H\"{o}rmann, J.~Hrubec, M.~Jeitler\cmsAuthorMark{1}, A.~K\"{o}nig, M.~Krammer\cmsAuthorMark{1}, I.~Kr\"{a}tschmer, D.~Liko, T.~Matsushita, I.~Mikulec, D.~Rabady, N.~Rad, B.~Rahbaran, H.~Rohringer, J.~Schieck\cmsAuthorMark{1}, J.~Strauss, W.~Treberer-Treberspurg, W.~Waltenberger, C.-E.~Wulz\cmsAuthorMark{1}
% ...
\vskip\cmsinstskip
% ...
1:~~Also at Vienna University of Technology, Vienna, Austria\\
% ...

\end{sloppypar}
\end{document}